\documentclass[12pt, a4paper]{article}

\usepackage[margin=1in]{geometry}

\usepackage[T1]{fontenc}  
\usepackage[utf8x]{inputenc}
\usepackage{amsmath}
\usepackage{pdfpages}
\usepackage{amsfonts}
\usepackage{amssymb}
\usepackage{multirow}
\usepackage{tikz}
\usepackage{graphicx}
\usetikzlibrary{positioning, decorations.pathreplacing, calc, arrows.meta}
\usepackage{bbold}
\usepackage{array}
\usepackage{subcaption}
\usepackage{float}

\newcommand{\R}{\mathbb R}

\newcommand{\dprod}[2]{\displaystyle\prod_{#1}^{#2}}
\newcommand{\dint}[2]{\displaystyle\int_{#1}^{#2}}

\usepackage[english]{babel} 
\usepackage{csquotes}

\usepackage[backend=biber,       % Utilise biber pour la compilation (recommandé)
    style=authoryear,    % Style de citation et de bibliographie de base
    maxcitenames=2,      % max 2 auteurs dans le texte, sinon "et al."
    maxbibnames=99,      % Liste tous les auteurs dans la bibliographie finale
    uniquename=false,    % Évite d'ajouter les initiales des prénoms dans le texte
    uniquelist=false,
    dashed=false,         % ne remplace pas les noms répétés par un tiret (---)
    url=false,        % Masque le champ URL
    eprint=false,      % Désactive l'affichage du champ eprint globalement
    giveninits=true,   % Active les initiales
    terseinits=true     % enlève le point après les initiales
]{biblatex}

\DeclareNameAlias{sortname}{family-given}
\DeclareNameAlias{default}{family-given}

\DeclareDelimFormat[{bibliography}]{nameyeardelim}{\addspace}

\DeclareDelimFormat{nameyeardelim}{\addspace} 
\DeclareFieldFormat[article]{title}{#1}

\DeclareFieldFormat{title}{#1}
\DeclareFieldFormat{journaltitle}{#1}

\renewbibmacro{in:}{}

\AtEveryBibitem{%
  \clearfield{note}
  \clearfield{month}
  \clearfield{day}
}

\providecommand{\keywords}[1]{\textbf{Keywords:} #1}
\title{
Using Temporal Replication and Partial Pooling to Improve Practical Identifiability in Presence-Only Poisson Point Process Models
}

\author{
  Camille Mottier$^{*\,1,2}$,
  Iris Lang$^{3}$,
  Justine Nicolas$^{3}$,
  Olivier Gimenez$^{2}$
}

\date{}

\begin{document}
\maketitle 

$^1$ Institut Montpelliérain Alexander Grothendieck (IMAG), France.\\
$^2$ Centre d'Écologie Fonctionnelle et Évolutive (CEFE),  France.\\
$^3$ Conservatoire d'espaces naturels (CEN) Occitanie, France.

\

$^{*}$ Corresponding author. E-mail: camille.mottier@umontpellier.fr

\vspace{1cm}

\begin{abstract}
Species distribution models based on thinned point processes are now commonly used in ecology to analyze presence-only data collected through citizen science, including in spatio-temporal frameworks. A key challenge in these models is the separation of the ecological process from the observation process, which raises identifiability issues. Here, we investigate the theoretical and practical identifiability of a thinned Poisson point process model. We show that, with a log link for ecological intensity and a logit link for observation probability, the model can be structurally identifiable under simple conditions, provided that observation probability varies spatially. We then evaluate the benefits of using temporal replication within a spatio-temporal framework to improve estimation when ecological intensity and observation probability are weakly separated. 
We quantify these benefits through simulations. The results highlight persistent practical identifiability problems affecting the ecological and observation intercepts, especially when observation probabilities are low. However, in a spatio-temporal model with partial pooling, these parameters are estimated more accurately than in a single-year model, even when only a short time period is considered. Temporal replication also improves estimation of parameters shared across years. Finally, we illustrate the use of such a spatio-temporal model through a study of coypu (\textit{Myocastor coypus}) distribution in Occitanie, France. Overall, our results show that temporal replication can partially alleviate practical identifiability problems in presence-only point process models when independent data sources are unavailable.
\end{abstract}
\keywords{Opportunistic data; Spatio-temporal modeling; Species distribution model.}

\vspace{1cm}

\section{Introduction}
In the current context of global change, invasive species are recognized as one of the main direct drivers of biodiversity loss (ref IPBES, \cite{roy_resume_2024}).
With rates of introduction reaching unprecedented levels, more than 37,000 established alien species are now recorded worldwide, including 3,500 considered invasive alien species, with impacts on habitats, ecosystems, as well as nature’s contributions to people.
Supporting management responses to their spread, including prevention, detection, eradication, containment and control, requires species distribution models (SDM), adapted to the particular challenges posed by invasive species. 
The recent and sometimes rapid emergence of alien species in new territories often implies the absence of systematic monitoring protocols and, consequently, of standardized data.  
For many species, opportunistic observations, including citizen-science observations, therefore constitute an important, and sometimes the main, source of distribution data, often providing broad spatial and temporal coverage. However, such data are typically presence-only and strongly affected by site accessibility, detectability, identification difficulty, and weather conditions. Thus, the spatial distribution of  observations  jointly reflects the actual distribution of the species and the observation process. Disentangling these two components is therefore a central challenge when inferring species distributions from opportunistic data.

Point processes provide a particularly suitable  framework for constructing SDM  based solely on presence-only data. 
\textcite{warton_poisson_2010} proposed an influential formulation in which species observations are treated as realization of an inhomogeneous Poisson point process whose intensity is related to environmental covariates. 
This model was subsequently extended to  account for the observation process, notably by \textcite{dorazio_accounting_2014},  using a thinning process governed by observation covariates. A key limitation, however, is that ecological intensity and observation effort can be difficult to distinguish, leading to identifiability problems, particularly when observation probabilities are uniformly low.
One common solution is to integrate presence-only observations with data collected through standardized protocols, such as presence-absence, count, or distance-sampling surveys. Because these additional data are affected by different observation processes, data integration can help overcome identifiability issues  (see for example \cite{dorazio_accounting_2014}, \cite{koshkina_integrated_2017}, \cite{farr_integrating_2021}, \cite{twining_integrating_2024}).  \textcite{fletcher_practical_2019}  review this increasingly common practice and provide guidelines its implementation. Yet standardized data are often unavailable, particularly for recently established invasive species. Opportunistic datasets, however, often provide observations over many years. By repeatedly observing the system under changing ecological and observation conditions, temporal replication may provide additional information to help disentangle these two processes when no independent data source is available.

Here, we assess this possibility using spatio-temporal thinned Poisson point process models fitted to presence-only data. We first examine theoretical identifiability (Section \ref{sec:modèle}) and then use simulations to assess when temporal replication and partial pooling improve the separation of ecological intensity from the observation process (Section \ref{sec:simu}). We finally illustrate the approach using 15 years of presence-only records of coypus (\textit{Myocastor coypus}) in Occitanie, southern France (Section \ref{sec:ragondins}).

\section{Model}\label{sec:modèle}

We model presence-only observations using a thinned inhomogeneous Poisson point process, in which the observed point pattern results jointly from an underlying ecological intensity and an observation process. We first define the spatial model and then extend it to repeated observations through time.

\subsection{Thinned inhomogeneous Poisson point process}

Let $D\subset\mathbb{R}^2$ denote the study region and let \[\mathbf{X}\sim PPP(\lambda)\] where $PPP(\lambda)$ denotes the Poisson point process with intensity $\lambda$. For any $B\subset D$, the expected number of points in $B$ is \[\Lambda(B)=\dint B{}\lambda(s)\,ds.\]

A thinning of $\mathbf{X}$ by a function $p: D\rightarrow [0,1]$ is the point process obtained by deciding independently for each point $s$ of the process $\mathbf X$ whether to retain it with probability $p(s)$ or remove it with probability $1-p(s)$. This defines a new Poisson point process denoted $\mathbf X_{\text{thin}}$, with intensity function  
$\lambda_\text{thin}=\lambda p$: \[\mathbf{X}_{\text{thin}}\sim PPP(\lambda p).\]

For general background on Poisson point process and thinning, see \textcite{moller_statistical_2004}.

\subsection{Species distribution model}

\begin{figure}[!h]
    \centering
    \input{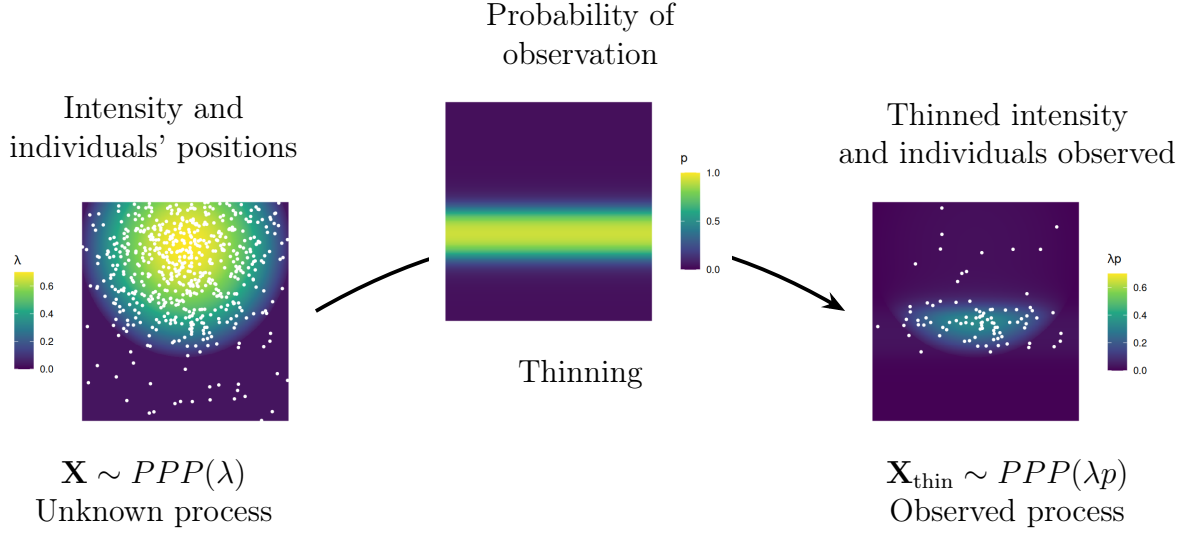}
    \caption{Species distribution model based on a thinned Poisson point process. }
    \label{fig:IPPP}
\end{figure}

In the ecological setting considered here, $\lambda(s)$ represents the spatial intensity of the species, whereas $p(s)$ represents the probability that an individual at location $s$ contributes an observation. This formulation extends point-process models for presence-only data \parencite{warton_poisson_2010} by explicitly representing observation bias through thinning (\cite{chakraborty_point_2011}, then \cite{fithian_finite-sample_2013} and \cite{dorazio_accounting_2014}).
The observation process may encompass both sample bias (sites less or more visited) and detection bias (difficulty to see and identify correctly the individuals). Presence-only records $s_{\rm obs}$ are therefore treated as a realization of the thinned process with intensity $\lambda p$ (Figure \ref{fig:IPPP}).
The ecological quantity of interest is $\lambda(s)$, which describes spatial variation in expected abundance per unit area. 
The Poisson formulation assumes independence among individual locations and among observation events conditional on the modeled intensities (in \ref{annexe:lgcp_2}, we examine a log-Gaussian Cox process extension allowing residual spatial dependence).

We model ecological intensity and observation probability as functions of distinct sets of covariates, using log and logit links as in \textcite{dorazio_accounting_2014}:
\[\forall s\in D,~\log(\lambda(s))=\beta_0+\beta x^t(s) ~ \text{ et }~\text{logit}(p(s))=\alpha_0+\alpha w^t(s),\]
where $x=(x_1,\ldots,x_m)$ and $w=(w_1,\ldots,w_q)$ are intensity and thinning covariates respectively. The intercept $\beta_0$ controls the overall scale of ecological intensity, whereas the slopes describe spatial variation associated with the ecological covariates.   
The model likelihood is then given by:
\[\begin{array}{ll}
\mathcal L(\alpha_0,\alpha, \beta_0,\beta ; s_{\rm obs}) &\hspace{-.3cm}= %\dfrac 1{n!}
\exp\left(-\dint D{} \lambda(s) p(s)\,ds\right)\dprod{i=1}{n_{\rm obs}}\lambda(s_i)p(s_i)
\end{array}\]

Because $\lambda(s)$ and $p(s)$ enter the likelihood only through their product, separating ecological intensity from the observation process is fundamentally an identifiability problem.

\subsection{Spatio-temporal extension}

To exploit temporal replication, we extend the model to discrete time periods $t\in T$, taken here to represent years. For each year, the observed point pattern is assumed to arise from 
\[\mathbf{X}_{\text{thin},t} \sim PPP(\lambda_t p_t)\] 
where $\lambda_t$ is an intensity function and $p_t$ is a thinning probability. 
Assuming conditional independence among years, the joint likelihood is:
\[\dprod{t\in T}{}\left(\exp\left(-\dint D{} \lambda_t(s) p_t(s)\,ds\right)\dprod{s\in {s_{{\rm obs},t}}}{}\lambda_t(s)p_t(s)\right).\]
We allow covariates to vary through time while sharing their effects across years:
\[
\forall t \in T,~\forall s\in D,~\log(\lambda_t(s))=\beta_{0,t} + \beta x^t(s,t)~\text{ and } ~ \text{logit}(p_t(s))=\alpha_{0,t}+\alpha w^t(s,t).
\]
The intercepts $\alpha_{0,t}$ and $\beta_{0,t}$ capture annual variation in ecological intensity and observation probability, respectively. Temporal structure can then be imposed on either set of intercepts, for example through constant effects, temporal trends, or random effects.

\subsection{Model identifiability}\label{sub-sec:identif}

\subsubsection{Structural identifiability}
Because only the product $\lambda(s)p(s)$ is observed, structural identifiability depends on the functional forms used for the ecological and observation processes.
With log links for both processes, the model is not identifiable because  \mbox{$\lambda(s)p(s)=\exp(\beta_0+\beta\,x^t(s)+\alpha_0+\alpha\, w^t(s))$} so the two intercepts enter the likelihood only through their sum. The same confounding occurs for coefficients associated with shared or linearly dependent covariates.
With a log link for $\lambda$ and a logit link for $p$, as used here, this symmetry is broken. Specifically, 
\[\log(\lambda(s)p(s)) = \beta_0+\beta x^t(s)+h(\alpha_0+\alpha w^t(s))\]
where $h$ denotes the function $\log\circ \text{logit}^{-1}$.

\

\textit{Case 1: constant detection probability}\\
The intercepts $(\alpha_0,\beta_0)$  are not identifiable when the observation probability $p$ is constant. Indeed, in the absence of a thinning covariate, for any $\alpha_0'\in\R$
\[\begin{array}{ll}
\log(\lambda(s)p(s))&=\beta_0+\beta x^t(s)+h(\alpha_0)\\
&=(\beta_0+h(\alpha_0)-h(\alpha_0'))+\beta x^t(s)+h(\alpha_0')
\end{array}\]
Hence, the pair of intercepts  $(\alpha_0', \beta_0')=(\alpha_0',\beta_0+h(\alpha_0)-h(\alpha_0'))$ yields the same likelihood as  $(\alpha_0, \beta_0)$. The intercepts are therefore not identifiable. Thus, identifiability requires the observation probability to vary spatially. 

\

\textit{Case 2: constant intensity and one covariate probability detection}\\
We consider the simple setting in which the intensity is constant $\lambda(s)=\exp(\beta_0)$ and the thinning probability depends on a single covariate $p(s)=\text{logit}^{-1}(\alpha_0+\alpha_1 w(s))$ with $\alpha_1\neq 0$ and $w$ non-constant. We assume that $w$ takes at least $n\geqslant 4$ distinct values $w_1<\ldots <w_n$ on $D$. Under these assumptions, the coefficients $(\alpha_0,\alpha_1,\beta_0)$ are identifiable from the thinned intensity. Indeed,  suppose that, for some coefficients $(\alpha_0',\alpha_1',\beta_0')$, we have:
\[\forall s\in D,~\beta_0+h(\alpha_0+\alpha_1 w(s))=\beta_0'+h(\alpha_0'+\alpha_1' w(s))\]
and  consider the function $f$ defined on $\R$ by $f(w)=\beta_0+h(\alpha_0+\alpha_1 w)-\beta_0'-h(\alpha_0'+\alpha_1' w)$. $f$ vanishes at $w_i$, for all $i$, and is differentiable on $\R$. Therefore, by Rolle's theorem (see for example \cite{strang}),  there exist $n-1$ points $c_1<\ldots <c_{n-1}$ such that, for all $i$, $f'(c_i)=0$, that is \mbox{${\alpha_1}/{(1+e^{\alpha_0+\alpha_1 c_i})}-{\alpha_1'}/(1+e^{\alpha_0'+\alpha_1'c_i})=0$.}
Since $\alpha_1$ is non-zero, $\alpha_1'$ must also be non-zero, and we therefore have  \mbox{$(1+e^{\alpha_0+\alpha_1 c_i})/{\alpha_1}-(1+e^{\alpha_0'+\alpha_1'c_i})/{\alpha_1'}=0$.} 
Thus, the  function $g$ defined by  $g(w)=(1+e^{\alpha_0+\alpha_1 w})/{\alpha_1}-(1+e^{\alpha_0'+\alpha_1'w})/{\alpha_1'}$ vanishes at $n-1$ points. Since it is differentiable on $\R$, Rolle's theorem implies that there exist $n-2$ points $d_1<\ldots <d_{n-2}$  such that, for all $i$, $g'(d_i)=0$, that is $e^{\alpha_0+\alpha_1 d_i}-e^{\alpha_0'+\alpha_1'd_i}=0$ or equivalently,  ${\alpha_0+\alpha_1 d_i}={\alpha_0'+\alpha_1'd_i}$. 
Since this linear equality holds at least two distinct points ($n\geqslant 4$), we conclude that $\alpha_0=\alpha_0'$ and $\alpha_1=\alpha_1'$. Then, using the initial equality, we obtain $\beta_0=\beta_0'$. Hence, the model is identifiable in this particular case. 

\

\textit{Other cases}\\
With multiple ecological and observation covariates, structural identifiability additionally depends on their relationships and cannot generally be established from link functions alone.  
In the spatio-temporal extension of the Poisson point process model, if, for each year $t$, the coefficients $(\alpha_{0,t},\alpha, \beta_{0,t},\beta)$ are identifiable from the thinned intensity $\lambda_tp_t$, then the  identifiability of the full model follows immediately.

\subsubsection{Practical identifiability}
Structural identifiability does not guarantee reliable estimation from finite and weakly informative data.  
A particularly important case occurs when observation probabilities are uniformly low \parencite{dorazio_accounting_2014}. 
Indeed, in the vicinity of 0, the logit function can be approximated by the log function, and we are once again in the case with two log links, with $\lambda(s)p(s)\simeq\exp(\beta_0+\beta x^t(s)+\alpha_0+\alpha w^t(s))$.
Practical identifiability can likewise deteriorate when ecological and observation models share the same covariates or include strongly correlated covariates (\cite{dorazio_accounting_2014}, \cite{koshkina_integrated_2017}).
Reliable separation of the two processes therefore requires informative and sufficiently distinct covariates, together with enough spatial variation in observation probability. In presence-only applications, however, the magnitude of observation probability is itself unknown. This motivates our simulation study, in which we ask whether temporal replication and partial pooling can improve estimation under conditions of poor practical identifiability.

\section{Simulations}\label{sec:simu}

We used simulations to assess practical identifiability of the spatio-temporal thinned Poisson point process model. Specifically, we asked whether: (1) temporal replication improves parameter estimation when observation probability is low; (2) any improvement extends from parameters shared across years to year-specific intercepts; and (3) temporal structure on the thinning intercepts is required to achieve such improvement.
 
\subsection{Simulated data}

The spatial domain was $D=[0 ; 50]^2$, partitioned into a $100\times 100$ raster grid with covariates constant within cells. We considered $T=\{1,\ldots,20\}$ years. 
We considered three simulation scenarios sharing the same ecological intensity process but differing in the temporal structure of the observation process: scenario 1 - constant detection intercept across years; scenario 2 - temporally varying detection intercept; scenario 3 - temporally varying detection intercept but more stable detection level.

\begin{figure}[!h]
    \centering
    \begin{subfigure}[b]{0.33\textwidth}
        \centering
        \includegraphics[width = \linewidth]{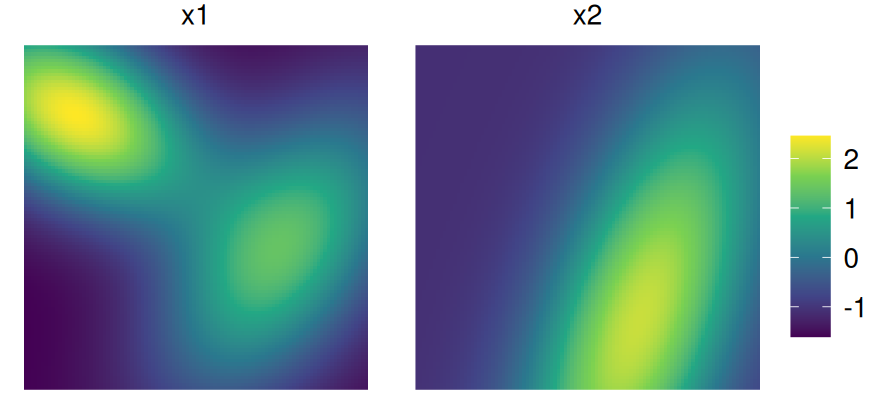}
        \caption{}
        \label{fig:x1_x2}
    \end{subfigure}%
    \hspace{.5cm}
    \begin{subfigure}[b]{0.6\textwidth}
        \centering
        \includegraphics[width=\linewidth]{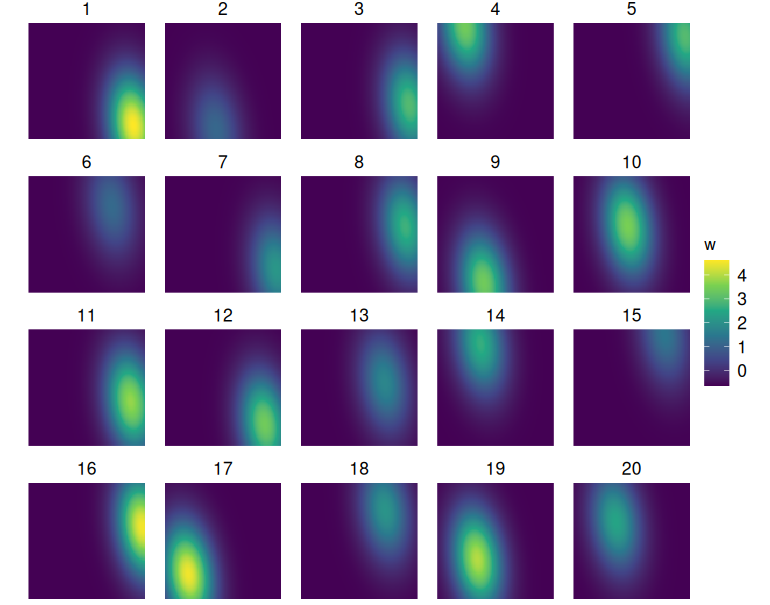}
        \caption{}
        \label{fig:w}
    \end{subfigure}
    \caption{Standardized covariates used in the simulations. (a) Intensity covariates $x_1$ et $x_2$ (all simulation scenarios). (b) thinning  covariates $w(\,.\,,t)$ for $t\in\{1,\ldots,20\}$ (scenario 1). }
\end{figure}

Two intensity covariates, $x_1$ and $x_2$ were held constant through time to represent  stable environmental conditions. Following \textcite{dorazio_accounting_2014}, they were generated from the density,  or the sum of densities, of multivariate normal distributions over $D$ and then standardized (Figure \ref{fig:x1_x2}). The intensity functions $\lambda_t$ were then defined as 
\[\forall t \in T,~\forall s\in D,~\log(\lambda_t(s))=\beta_{0,t}+2\,x_1(s) + x_2(s)\]
where $\beta_{0,t}$ was drawn from a $\mathcal N(-1,0.2^2)$.

Regarding the thinning process, we used a single covariate time-varying covariate $w(s,t)$ representing observer distribution. Each annual surface was generated a multivariate normal density distribution with fixed covariance and a mean drawn independently each year from a uniform distribution over $D$, producing spatially heterogeneous thinning pattern that varied across years.
Annual variation in this surface was controlled by an attenuation coefficient drawn from $\mathcal U(0.3,1)$ in scenarios 1 and 2, and from $\mathcal U(0.8,1)$ in scenario 3, resulting in stronger temporal variation in observer distribution in the first two scenarios. The covariate was then standardized across all years (Figure \ref{fig:w}). 
The thinning functions $p_t$ were defined as
\[\forall t \in T,~\forall s\in D,~\text{logit}(p_t(s))=\alpha_{0,t}+0.5\, w(s,t)\]
where $\alpha_{0,t}=-3.5$ in scenario 1 and $\alpha_{0,t} \sim \mathcal N(-3.5, 0.2^2)$ in scenarios 2 and 3.

Across all scenarios, the correlation between the covariates is below 0.83 and 
expected sample sizes were of similar magnitude (expected annual number of observed points: 106–399 in scenario 1, 150–684 in scenario 2, and 154–568 in scenario 3), allowing meaningful comparison of estimation performance.
For each scenario, the year with the highest observation probability was relabeled as year 1, providing the most favorable single-year benchmark (highest probability: 0.23, 0.35, and 0.18 in scenarios 1, 2, and 3, respectively). 
For each $t\in T$, 500 realizations of a Poisson point process with thinned intensity $\lambda_tp_t$ were simulated using the  \texttt{rpoispp} function from the  spatstat package \parencite{baddeley_spatial_2016}.

\subsection{Fitting}

For each of the three described above and each of the 500 simulated datasets, we fitted the spatio-temporal thinned Poisson point process using 1, 5, 10, 15 or 20 years of data. All fits included year 1, thereby comparing the most favorable single-year analysis with progressively longer temporal series. 
The intensity intercepts $\beta_{0,t}$ were always assumed independent across years, while the thinning intercepts  $\alpha_{0,t}$ were modeled as either constant (model $\mathcal M_\text{cst}$, scenario 1 only), independent across years (model $\mathcal M_\text{ind}$), or as year-specific random effects (model $\mathcal M_\text{eff}$, with partial pooling across years):
\[\log(\lambda_t(s)) = \beta_{0,t}+\beta_1\, x_1(s)+\beta_2\, x_{2}(s), \text{ with independent $\beta_{0,t}$ }\]
\[\text{logit}(p_t(s)) = \alpha_{0,t}+\alpha_1\, w(s,t), 
\text{ where } \alpha_{0,t}= \left\{\begin{array}{l}
\alpha_{0} \text{  in }\mathcal M_\text{cst} \text{ (scenario 1 only)}\\ 
\text{independent across years in }\mathcal M_\text{ind} \\
\mathcal N(\mu, \sigma^2) \text{ in }\mathcal M_\text{eff} 
\end{array}\right.\]

We fitted models in a Bayesian framework using INLA \parencite{rue_approximate_2009} through the \texttt{lgcp} function of the \textit{inlabru} package \parencite{bachl_inlabru_2019}, with integration points constructed from a regular triangular mesh over $D$ (maximum edge length 2).
All model parameters were assigned Gaussian priors $\mathcal N(0,1000)$, except for  the variance parameter $\sigma^2$, for which the  prior was defined by $\sigma^2=e^{-\theta}$, where $\theta$ follows a log-gamma distribution with parameters $(1, 5.10^{-5})$.

\subsection{Simulation results}

We assessed using relative bias, quadratic risk, mean $95\%$ credible-interval width, and coverage probability, based on posterior means across simulations. Three runs with questionable convergence were excluded in scenarios 1 and 3; none were excluded in scenario 2.

\subsubsection{Temporal replication improves estimation }

\begin{figure}[!htbp]
    \begin{subfigure}{\textwidth}
    \centering
        \includegraphics[width=0.7\linewidth]{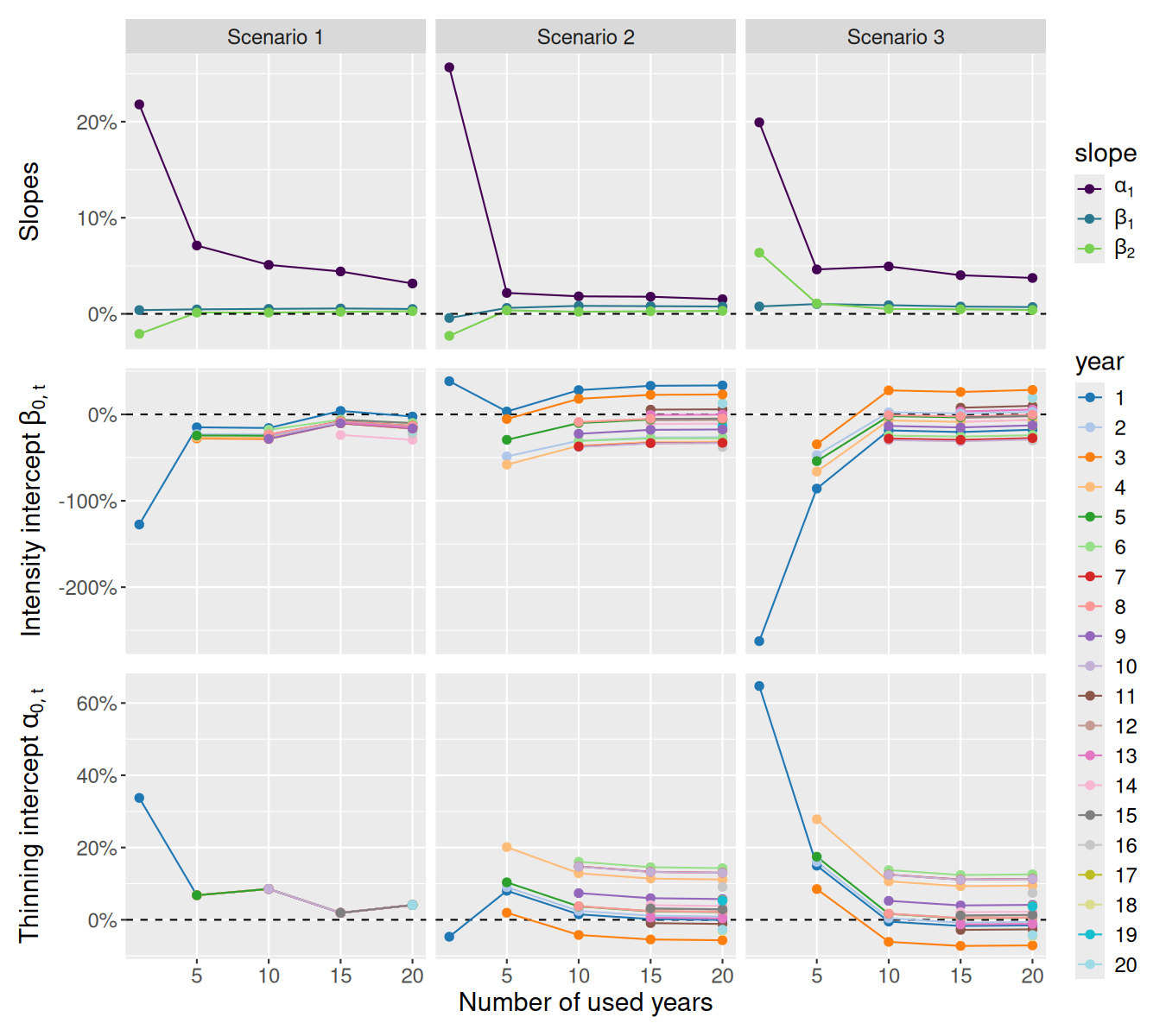}
    \caption{Model $\mathcal M_\text{eff}$}
    \label{fig:bias_ind_eff}
    \end{subfigure}
    \vfill
    \begin{subfigure}{\textwidth}
    \centering
        \includegraphics[width=0.7\linewidth]{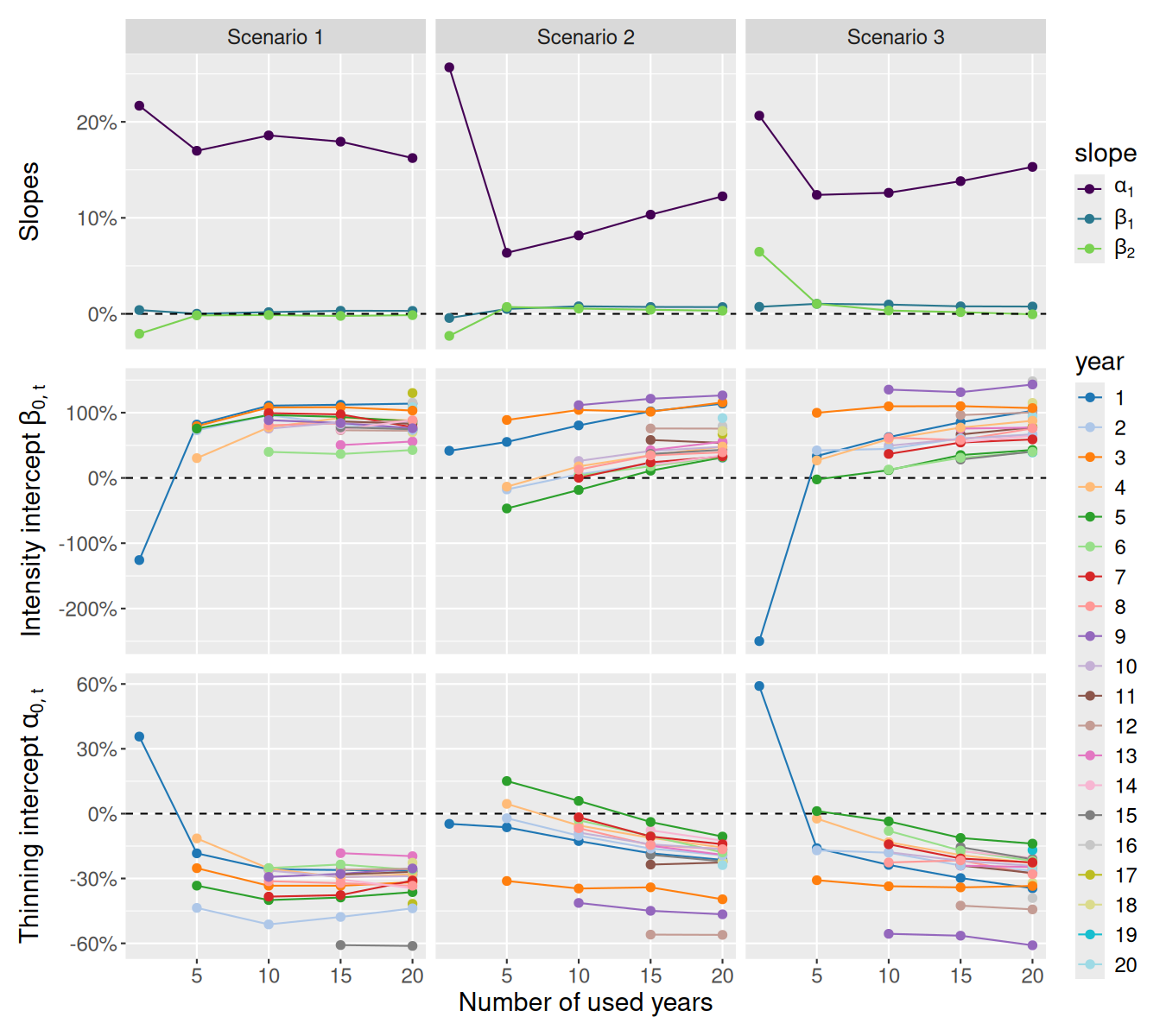}
    \caption{Model $\mathcal M_\text{ind}$}
    \label{fig:bias_ind_ind}
    \end{subfigure}
    \caption{Estimated bias for each parameter and simulation scenario, for models including 1, 5, 10, 15, or 20 years of data. (a) Model $\mathcal M_\text{eff}$ assumes intensity intercepts $\beta_{0,t}$ that are independent across years and thinning intercepts $\alpha_{0,t}$ modeled as year-specific random effect. (b) Model $\mathcal M_\text{ind}$ assumes intensity intercepts $\beta_{0,t}$ and thinning intercepts $\alpha_{0,t}$ that are independent across years.}
\end{figure}

We first considered $\mathcal M_\text{eff}$, in which thinning intercepts are partially pooled across years. Relative bias is shown in Figure \ref{fig:bias_ind_eff} and the remaining performance metrics in Figure \ref{fig:indic_ind_eff}. 

As expected under low observation probabilities, intercepts were substantially harder to estimate than slopes, with ecological and thinning intercepts showing biases of opposite sign \ref{fig:bias_ind_eff}. 
Temporal replication improved estimation of the shared slopes $\beta_1$, $\beta_2$, $\alpha_1$, with bias generally decreasing as additional years were included. 
Importantly, temporal replication also improved estimation of year-specific intercepts in scenarios 1 and 3, despite these parameters not being shared across years. Most of the improvement occurred within the first 5-10 years, after which gains were limited. Scenario 2 showed weaker improvement, partly because its single-year benchmark was already comparatively well estimated. 
Quadratic risk and credible-interval width also decreased as more years were included (Figure \ref{fig:indic_ind_eff}), indicating gains in precision even when bias changed little. Intercepts nevertheless remained less precisely estimated than slopes. 
Coverage of slope parameters remained above $90\%$, whereas coverage for intercepts was poorer (Figure \ref{fig:indic_ind_eff}). Moving from one to five years substantially improved intercept coverage, although coverage sometimes declined again as credible intervals became narrower with longer time series.
Overall, temporal replication under $\mathcal M_\text{eff}$ improved both shared slopes and year-specific intercepts, with substantial gains already achieved using five years of data, although intercept estimation remained comparatively difficult.

\subsubsection{Partial pooling improves intercept estimation}

\begin{figure}[!htbp]
\begin{subfigure}{\textwidth}
    \centering
    \includegraphics[width=0.67\linewidth]{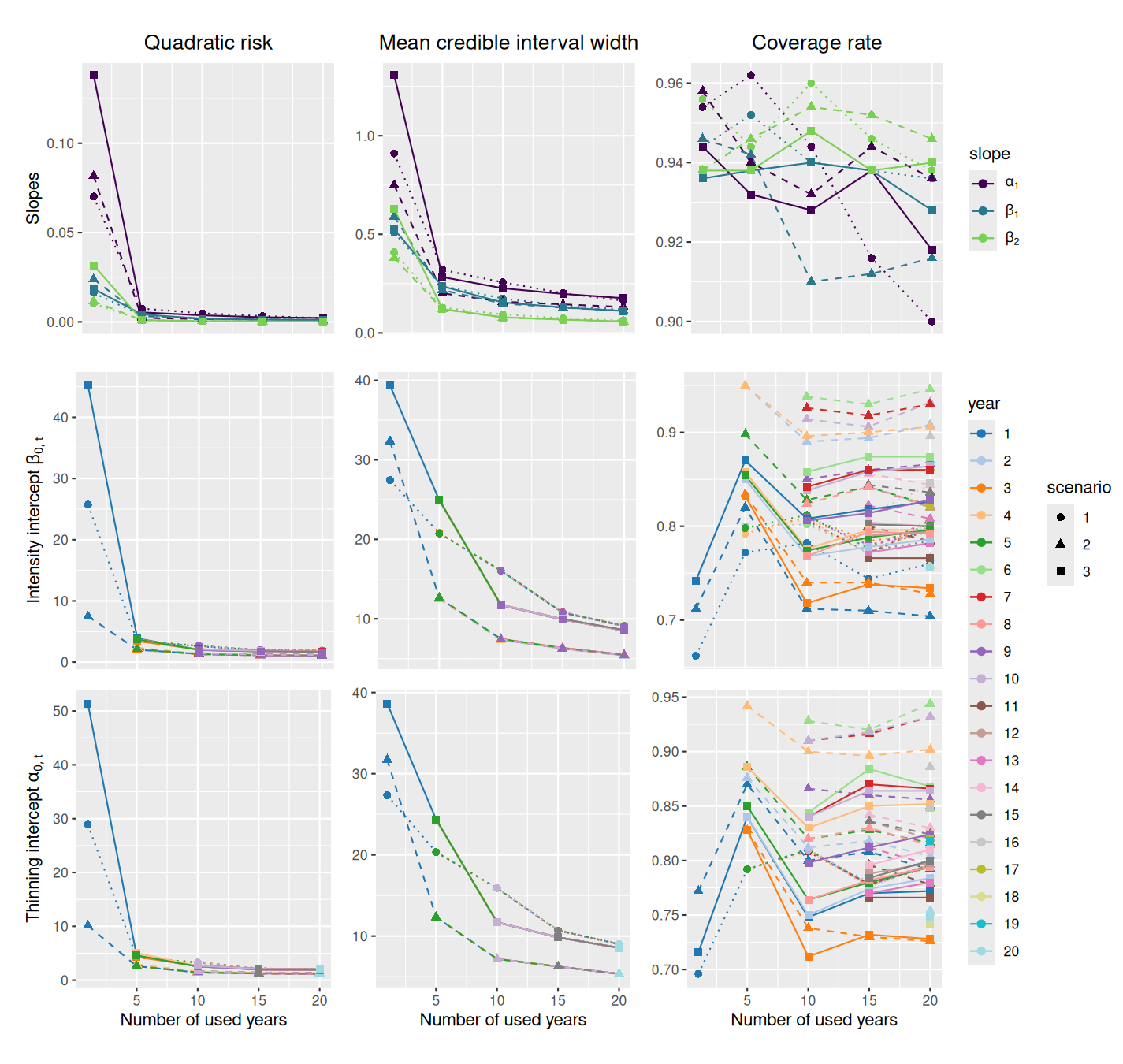}
    \caption{Model $\mathcal M_\text{eff}$}
    \label{fig:indic_ind_eff}
\end{subfigure}
\vfill
\begin{subfigure}{\textwidth}
    \centering
    \includegraphics[width=0.67\linewidth]{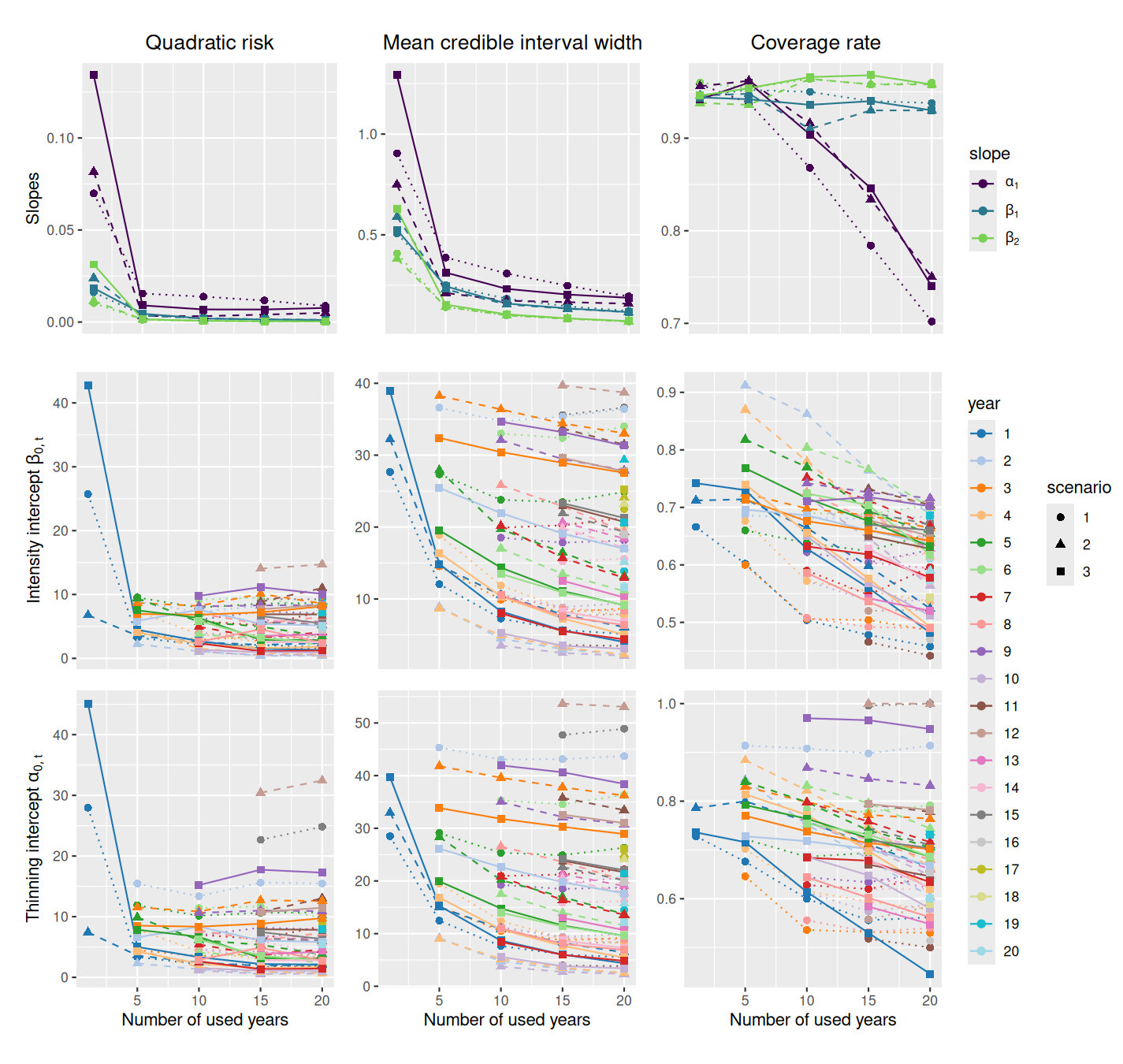}
    \caption{Model $\mathcal M_\text{ind}$ }
    \label{fig:indic_ind_ind}
\end{subfigure}
\caption{Quadratic risk, mean credible interval width, and coverage rate for each parameter and simulation scenario, for models including 1, 5, 10, 15, or 20 years of data. (a) Model $\mathcal M_\text{eff}$ assumes intensity intercepts $\beta_{0,t}$ that are independent across years and thinning intercepts $\alpha_{0,t}$ modeled as year-specific random effect. (b) Model $\mathcal M_\text{ind}$ assumes intensity intercepts $\beta_{0,t}$ and thinning intercepts $\alpha_{0,t}$ that are independent across years.}
\end{figure}

We next removed partial pooling by fitting $\mathcal M_\text{ind}$ with independent thinning intercepts across years (Figures \ref{fig:bias_ind_ind} and \ref{fig:indic_ind_ind}).

The ecological slopes $\beta_1$ and $\beta_2$ remained well estimated, but gains for the thinning slope $\alpha_1$ were weaker than under $\mathcal M_{eff}$. Although its quadratic risk and credible-interval width decreased with additional years, bias changed little and coverage declined to approximately $70\%$ for the longest time series.  
Intercept estimation deteriorated much more strongly without partial pooling, with relative biases exceeding $100\%$ for some parameters and coverage below $50\%$. Temporal replication alone therefore provided only limited information about year-specific intercepts when these were left unconstrained.
Overall, sharing slope parameters across years was insufficient to resolve practical confounding of the intercepts. Partial pooling of the thinning intercepts provided the additional structure needed to substantially improve their estimation.

\

Results for $\mathcal M_\text{cst}$ in scenario 1 were very similar to those for $\mathcal M_\text{eff}$ and are provided in \ref{annexe:scen1}.

\section{Case study}\label{sec:ragondins}

We applied the spatio-temporal thinned Poisson point  process to a real presence-only dataset to assess whether the temporal structure that improved practical identifiability in simulations also helps separate ecological intensity from the observation process in practice.

\subsection{Data and covariates}

\begin{figure}
    \centering
    \includegraphics[width=.8\linewidth]{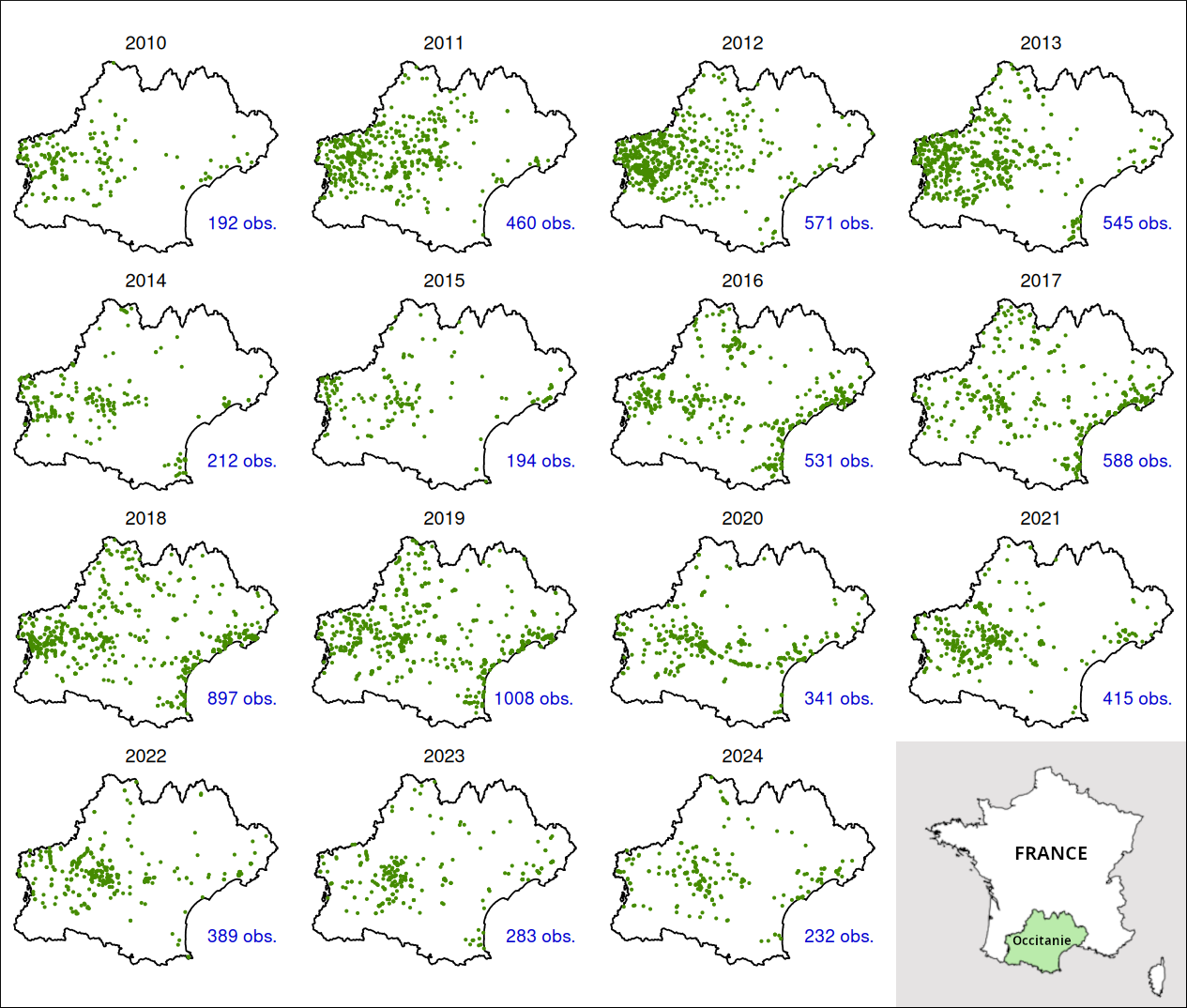}
    \caption{Coypu observation data collected between 2010 and 2024 in Occitanie, obtained from the SINP. In blue, the total number of distinct locations recorded per year.}
    \label{fig:données}
\end{figure}

The case study focuses the spatial distribution of coypus (\textit{Myocastor coypus})  in the Occitanie of southern France. The data were obtained from "Système d'information de l'inventaire du patrimoine naturel" \parencite{sinp}  and span a 15-year period, from 2010 to 2024 (see Figure \ref{fig:données}).
The dataset consists primarily of presence-only records subsequently validated by experts, and therefore combines substantial temporal replication with strongly heterogeneous observation effort across both space and time. To avoid duplicate records, we retained at most one observation per spatial location and year.

\begin{figure}[ht]
    \centering
    \includegraphics[width=0.24\linewidth]{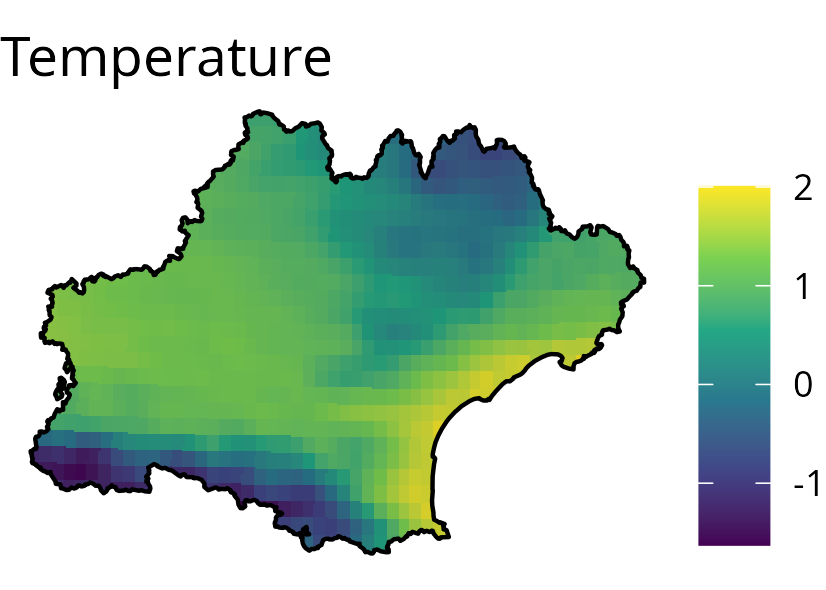}
    \includegraphics[width=0.24\linewidth]{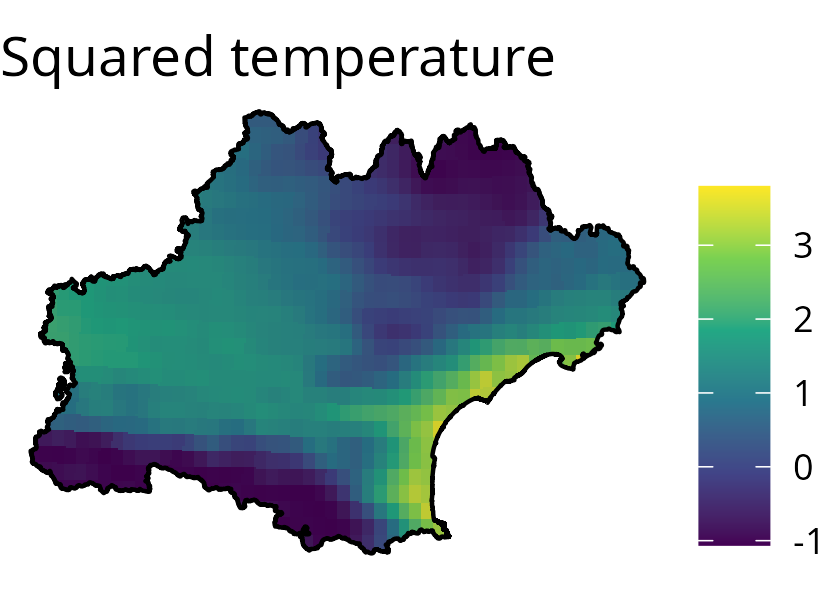}
    \includegraphics[width=0.24\linewidth]{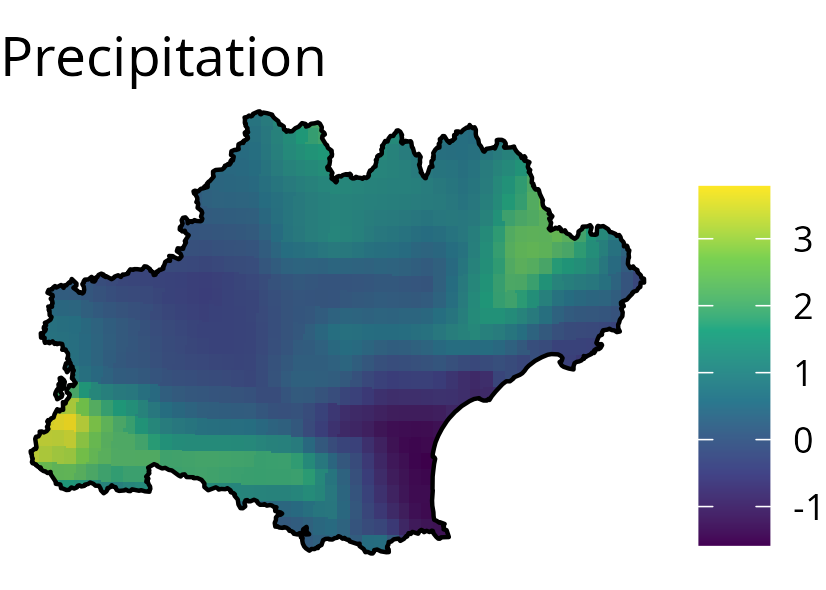}
    \includegraphics[width=0.24\linewidth]{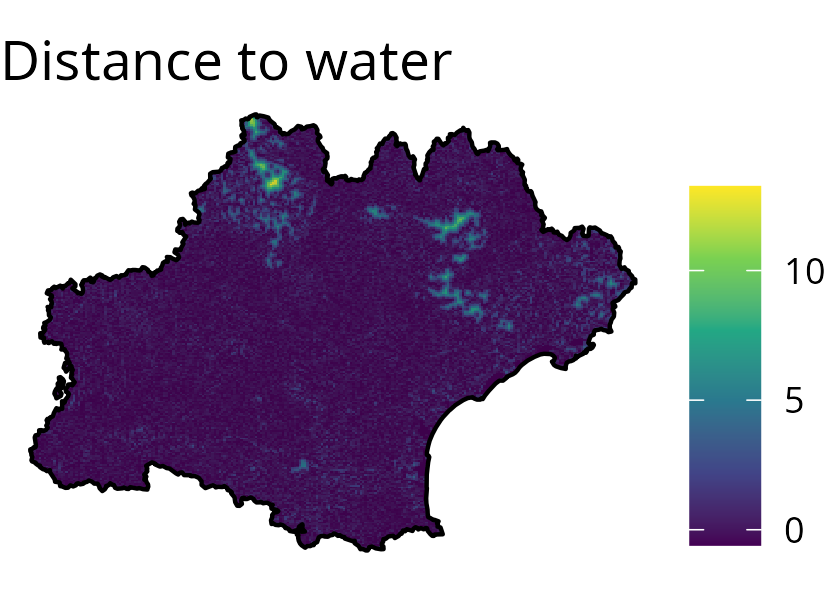}
    \includegraphics[width=0.24\linewidth]{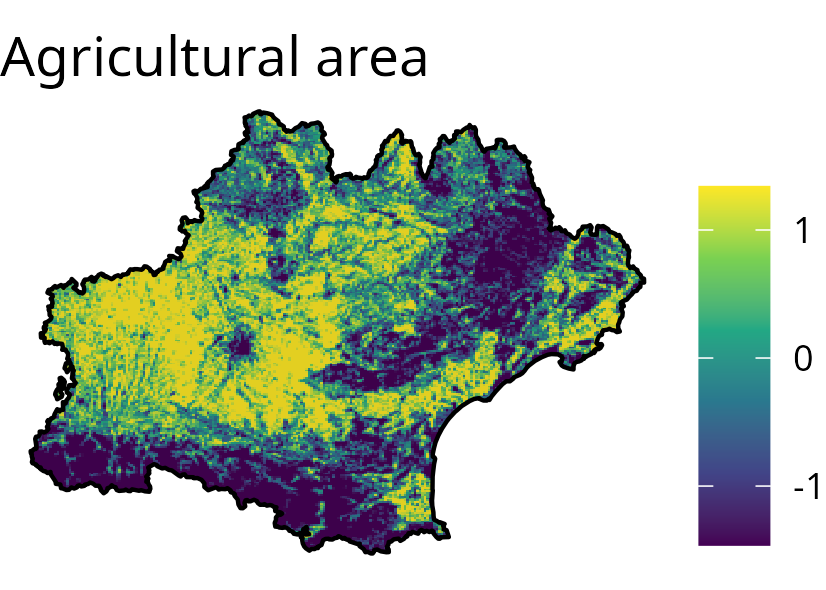}
    \includegraphics[width=0.24\linewidth]{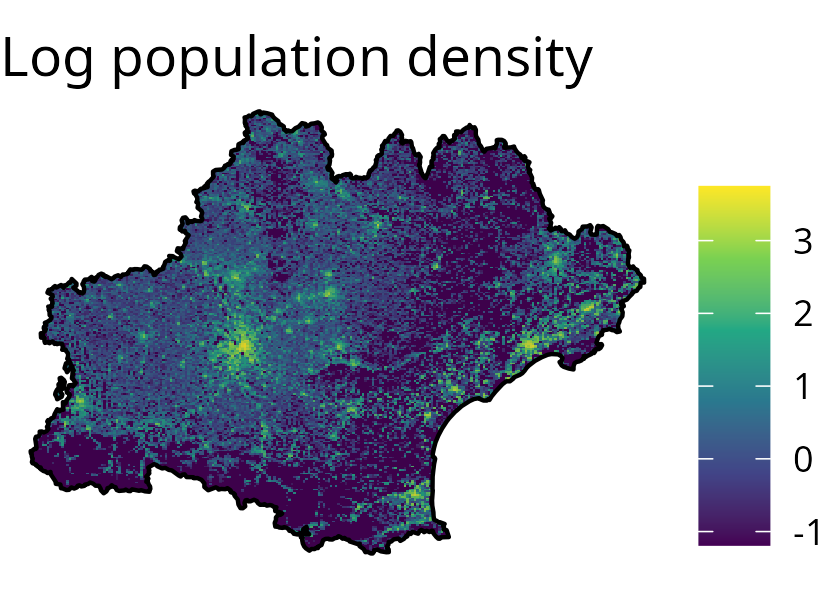}
    \includegraphics[width=0.32\linewidth]{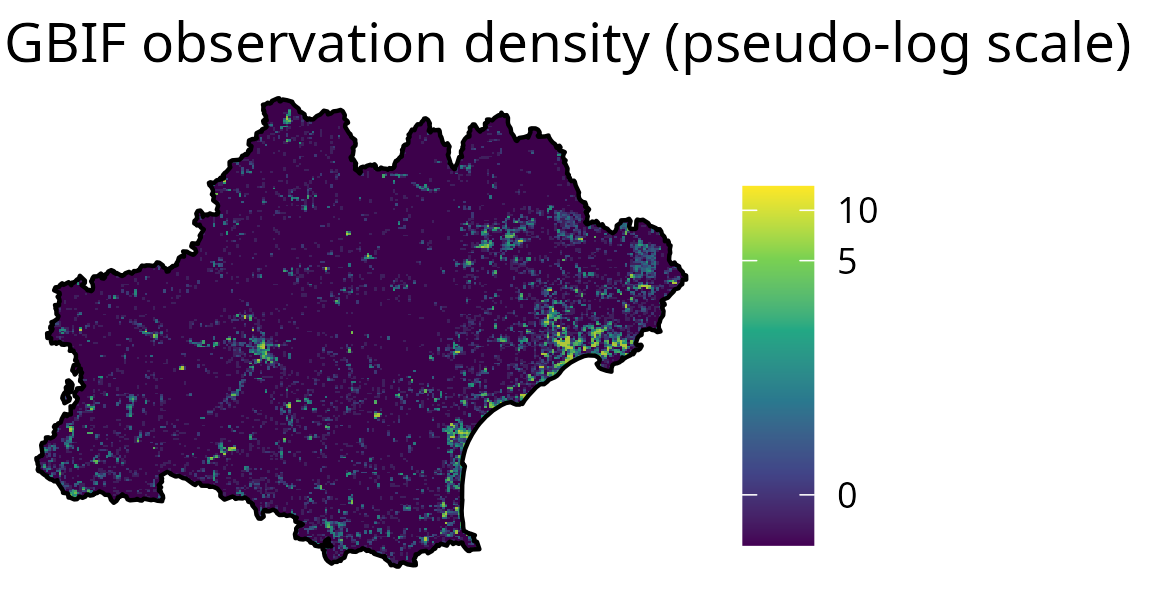}
    \caption{Intensity and thinning covariates used in the analysis for 2024, shown on a $250\times 250$ grid.}
    \label{fig:covar}
\end{figure}

Model covariates were selected based on previous studies of coypu distribution in Europe (\cite{hilts_climate_2019}, \cite{jarnevich_evaluating_2017}, \cite{schertler_potential_2020}, and \cite{vaissi_climatic_2023}). We considered six covariates related to temperature, precipitation, distance to water, agricultural land, and human population density. Their definitions and data sources are summarized in Table \ref{tab:covar}. All covariates were standardized. For temporally varying covariates, the same scaling was applied across years. The standardized covariates for 2024 are shown in Figure \ref{fig:covar}.

Thinning was modeled using a covariate representing observer presence, defined as the density of multi-taxon opportunistic records within a 1-km radius (Table \ref{tab:covar}), following \textcite{twining_integrating_2024}.
The data were obtained from the Global Biodiversity Information Facility (GBIF). We extracted bird, mammal, amphibian and reptile records, capping sampling at 2,000 observations per taxon and month, and removed duplicates by retaining only one observation per location, day and observer. This yielded $19,402-35,496$ records per year. 
Local observation density was capped at 50 records to limit the influence of heavily sampled locations; sensitivity to this threshold is examined in \ref{annexe:sensib}.
The resulting thinning covariate was standardized across the full study period and is illustrated  for 2024 in Figure \ref{fig:covar}.

\begin{table}[!ht]
    \centering
    \renewcommand{\arraystretch}{1.2}

    \begin{tabular}{|>{\raggedright}p{4.5cm}|p{8cm}|p{1.8cm}|}
    \hline
   \centering Covariate   & \centering Database 
   & Temporal resolution \\\hline
    Annual minimum of mean monthly temperatures  ($x_{temp}$) & ERA5-Land \parencite{copernicus_climate_change_service_era5-land_2019} 
    & Annual \\\hline
    Annual minimum of mean monthly squared temperatures ($x_{temp^2}$) & ERA5-Land \parencite{copernicus_climate_change_service_era5-land_2019}
    & Annual \\\hline
    Annual cumulative precipitation ($x_{prec}$) &  ERA5-Land \parencite{copernicus_climate_change_service_era5-land_2019} 
    & Annual \\\hline
    Distance to the nearest surface water ($x_{water}$)& France: BD TOPO$^{\mbox{\scriptsize{\textregistered}}}$  \parencite{bdtopo} \par 
    Spain (10-km buffer): \par hydroRIVERS \parencite{hydrorivers}  and hydroLAKES \parencite{hydrolakes}
    & Constant \\\hline
    Proportion of agricultural land within a 1-km radius ($x_{agri}$)& CORINE Landcover 2012 \parencite{corine}
    & Constant \\\hline
    Log-transformed population density ($x_{pop}$) & Census grid Eurostat 2021 \parencite{eurostat}
    & Constant\\\hline
    Multi-species observation density within a 1-km radius ($w_{gbif})$ & Global Biodiversity Information Facility \parencite{gbif} & Annual \\\hline
    
    \end{tabular}
    \caption{Intensity and thinning covariates used in the analysis and their data sources. The last column indicates whether the covariates remained constant over the 2010-2024 period or were updated annually.}
    \label{tab:covar}
\end{table}

\subsection{Fitting and results}

All models  were fitted using the \texttt{lgcp} function from the \textit{inlabru} package \parencite{bachl_inlabru_2019}. Integration points were generated using the  \texttt{fm\_int} function from the \text{fm\_mesher} package and a triangular mesh with edges lengths  ranging from 2.5 km and 20 km, incorporating both the coypu and GBIF data used (Figure \ref{fig:mesh}).

\

We fitted the spatio-temporal thinned Poisson point process over the full 2010–2024 period, using the same general structure as in the simulation study:
\[\log(\lambda_t(s))=\beta_{0,t}
+\beta_{1}\,x_{temp}(s,t)
+\beta_{2}\,x_{temp^2}(s,t)
+\beta_{3}\,x_{prec}(s,t)
+\beta_{4}\,x_{water}(s)
+\beta_{5}\,x_{agri}(s)
+\beta_{6}\,x_{pop}(s),\]
\[\text{logit}(p_t(s))= \alpha_{0,t}+\alpha_1\,w_{gbif}(s,t).\]

We compared alternative temporal structures for both sets of intercepts. Observation intercepts $\alpha_{0,t}$ were modeled as constant, independent across years, or as year-specific random effects; ecological intercepts $\beta_{0,t}$ were modeled as constant, linear in time, independent across years, or as year-specific random effects. Each model was fitted with either a linear or quadratic temperature effect.
All parameters were assigned Gaussian priors $\mathcal N(0,100)$, or $\mathcal N(0,1000)$ for the independent intercepts, except for the random-effect variance parameter, parametrized as $\sigma=e^{-\theta}$, with $\theta$ following a log-normal distribution $(1,5.10^{-5})$.

\begin{figure}[ht]
    \centering
    \includegraphics[width=0.9\linewidth]{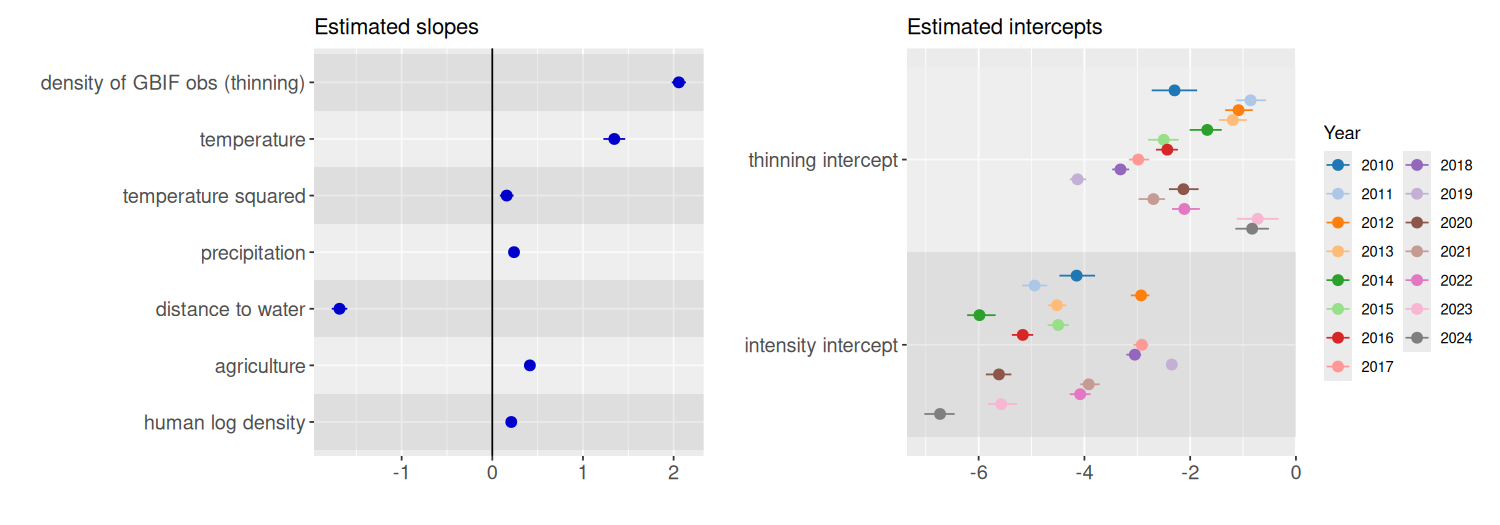}
    \caption{Estimated intercepts (posterior means) and associated $95\%$ credible intervals for the spatio-temporal thinned Poisson point process model fitted with annual random-effect  intercepts and quadratic temperature effect.}
    \label{fig:coeff_eff_eff}
\end{figure}

Models were compared using WAIC. 
Models with independent or random-effect ecological and observation intercepts clearly outperformed more constrained temporal structures, with WAIC values ranging from 62,460 to 62,482 for the former models and from 63,353 to 66,229 for the latter. Models including a quadratic temperature effect generally had lower WAIC values. The four best-supported models differed little in WAIC and produced nearly identical estimates. We therefore present results for the model with random effects for both intercepts and a quadratic temperature effect, which provided the best fit according to WAIC; results for the alternative models are given in \ref{annexe_comparaison}. 
Figure \ref{fig:coeff_eff_eff} shows the estimated parameters and their associated $95\%$ credible intervals. The strongest associations were with temperature and distance to water: coypu intensity increased with temperature and decreased with distance from water, consistent with the species' ecology. The slope associated with the thinning covariate was positive, indicating higher coypu observation probability in areas with greater multi-taxon recording activity. In contrast, yearly intercepts were much less precisely estimated, with wide credible intervals and substantial temporal variation, consistent with the identifiability issues highlighted by the simulations. This variation may partly reflect incomplete characterization of observation effort by the GBIF-based thinning covariate.

\begin{figure}[ht]
    \centering
    \includegraphics[width=0.5\linewidth]{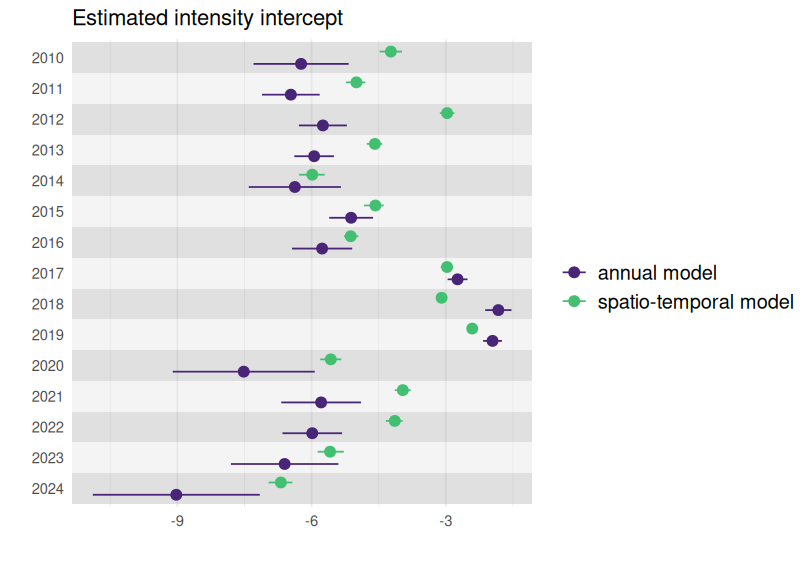}
    \caption{Comparison of  intensity intercepts from the spatio-temporal thinned Poisson point process model with annual random-effect intercepts and quadratic temperature effect and from separate single-year thinned Poisson point process models. }
    \label{fig:comp_annual_spat_temp}
\end{figure}

Compared with separate single-year fits, the spatio-temporal model produced substantially narrower credible intervals for the ecological intercepts (Figure \ref{fig:comp_annual_spat_temp}), indicating stronger information sharing across years. However, estimates from the two approaches often differed markedly, and their credible intervals rarely overlapped. This discrepancy reinforces the need for caution when interpreting yearly intercepts and absolute intensity.

\begin{figure}[ht]
    \centering
    \includegraphics[width=0.4\linewidth]{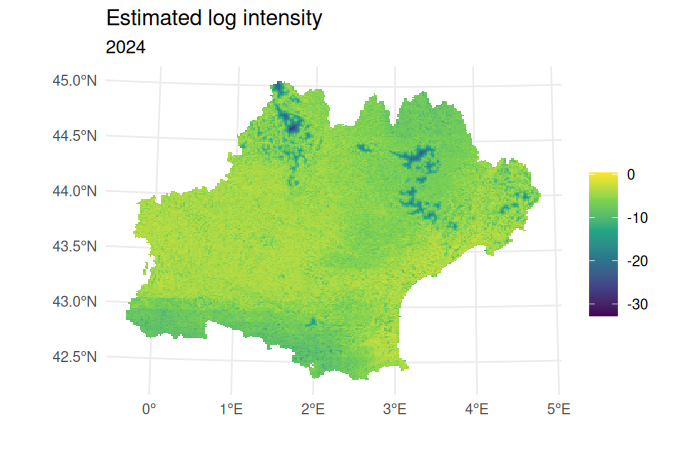}
    \includegraphics[width=0.4\linewidth]{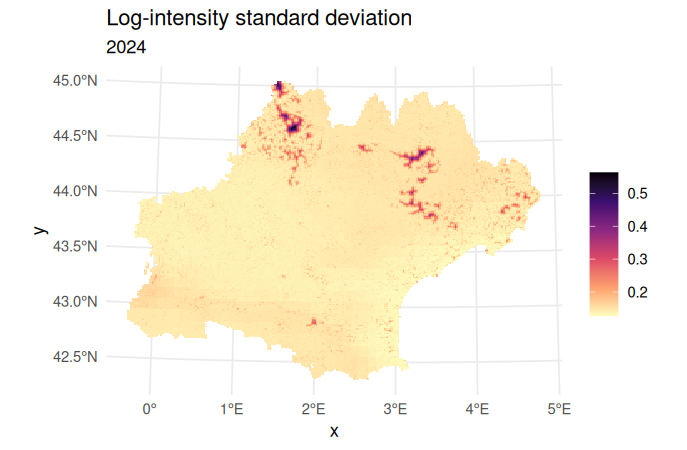}
    \caption{Log-intensity estimated (posterior mean) and associated standard deviation for 2024 obtained from the spatio-temporal thinned Poisson point process model with annual random-effect intercepts and quadratic temperature effect.}
    \label{fig:cartes_ppp}
\end{figure}

Finally, Figure \ref{fig:cartes_ppp} shows the estimated log-intensity for 2024 together with its associated standard deviation. Maps for all years are provided in \ref{annexe:carte_periode}.

\section{Discussion}\label{sec:discussion}

In this study, we investigated the conditions under which the ecological and  observation processes of a thinned Poisson point process can be disentangled from presence-only data. Our results highlight an important distinction between theoretical and practical identifiability. 
We showed that combining a log link for ecological intensity with a logit link for thinning can yield theoretical identifiability under simple conditions, provided that observation probabilities varies spatially. We also showed that  low observation probability can lead to strong practical confounding between ecological and observation intercepts. 

Our simulations show that temporal replication can substantially improve practical identifiability when incorporated within a spatio-temporal model. This approach improves the estimation of shared parameters, and consequently of the relative intensity (i.e., up to a multiplicative constant), and can also improve estimation of year-specific intercepts under low sampling effort. However, temporal replication alone does not resolve their confounding; partial pooling of the observation intercepts is required to borrow information across years. Even then, intercepts remain substantially less precisely estimated than slopes. Compared with fitting separate single-year models, most of the gain is already achieved with five years of data, with relatively little additional imrovement thereafter. This suggests that temporal structure can be informative even for datasets covering relatively few years. A natural extension would be to assess how these conclusions depend on the choice of link function and model specification.

In the case study on the spatial distribution of coypus, models with independent and random intercepts produced very similar results, unlike in the simulations where partial pooling clearly improved estimation. This may indicate either that the case-study data already contain sufficient information for the shared parameters, or that the random-effects structure adds little information beyond that contained in the observations. In either case, the spatio-temporal formulation improves estimation of parameters shared across years and therefore of relative ecological intensity. Inference on relative intensity nevertheless depends on how well the chosen covariates represent the ecological process. More flexible covariate structures could improve predictive performance, but this was not the focus of the present study.  
In light of the identifiability study presented above, some uncertainty remains regarding the estimated intercepts and, consequently, the absolute intensity. In particular, the large year-to-year variations in the intercepts, with generally opposite patterns between the intensity and the thinning intercepts, are consistent with the practical confounding identified in our theoretical and simulation analyses.
This variability may partly reflect an incomplete characterization of observation effort. Constructing suitable effort covariates from opportunistic data  is difficult because observer activity is only indirectly measured and may itself be spatially associated with ecological covariates. Thus, achieving separation of the two processes requires not only appropriate link functions and a sufficient data, but also informative covariates for both ecological intensity and observational effort. 

A thinned log-Gaussian Cox process (LGCP) offers one possible extension by accounting for residual spatial variation and spatial dependence through a Gaussian random field (see, for example, \cite{pant_spatio-temporal_2025}), and their implementation has been greatly facilitated by the INLA inference framework  \parencite{rue_approximate_2009}.
However, adding such flexibility does not remove the identifiability problem: unexplained spatial variation must still be attributed to either the ecological or the observation process.
In our case study, the Poisson and LGCP formulations produced very similar results \ref{annexe:lgcp_2}, suggesting that additional spatial structure alone does not resolve this issue.

Regardless of whether a Poisson or an LGCP model is used, model validation remains challenging. Without independent information, goodness-of-fit diagnostics necessarily assess the observed, thinned  process rather than the underlying ecological process. For example, the inhomogeneous Ripley's $K$ function \parencite{baddeley_non_2000} used in \ref{annexe:valid} evaluates fit to the observed point pattern but cannot determine whether ecological intensity and observation effort have been correctly separated. 

Opportunistic records, including citizen-science records, often provide the main, and sometimes the only, information available on the spatial distribution of invasive alien species. When standardized data are unavailable, integration with independent data sources is therefore not possible.  Our results show that temporal replication offers a useful alternative source of information: by combining repeated observations through a spatio-temporal model, presence-only datasets can yield better estimates of relative ecological intensity and, under suitable hierarchical structure, partially alleviate practical identifiability problems. Temporal replication does not replace independent data, but it represents an underused source of information that can substantially strengthen inference from opportunistic biodiversity records. 

\vspace{1cm}

\textbf{Availability of data and materials} \\
All relevant data and code used to produce the simulation analyses are available at \url{https://anonymous.4open.science/r/spat-temp_LGCP-962F/}\\
The coypu observation dataset used during the current case study is not publicly available but is available from \url{https://sinp-occitanie.fr/atlas/static_pages/diffusion} on request. The code to fit the model is available at \url{https://anonymous.4open.science/r/spat-temp_LGCP-962F/}

\vspace{1cm}

\textbf{Acknowledgements}\\
This work received support from: the French government, managed by the Agence Nationale de la Recherche (ANR) as part of France 2030, and from the European Union’s NextGenerationEU programme under the reference « ANR-21-EXES-0005 » ; from the Occitanie Region;and from the ExposUM Institute of the University of Montpellier.

\appendix

\renewcommand{\thesection}{Appendix \Alph{section}}
\renewcommand{\thesubsection}{\Alph{section}.\arabic{subsection}}

\section{Results for the $\mathcal M_\text{cst}$ model under simulation scenario 1}\label{annexe:scen1}

\begin{figure}[ht]
    \centering
    \includegraphics[width=\linewidth]{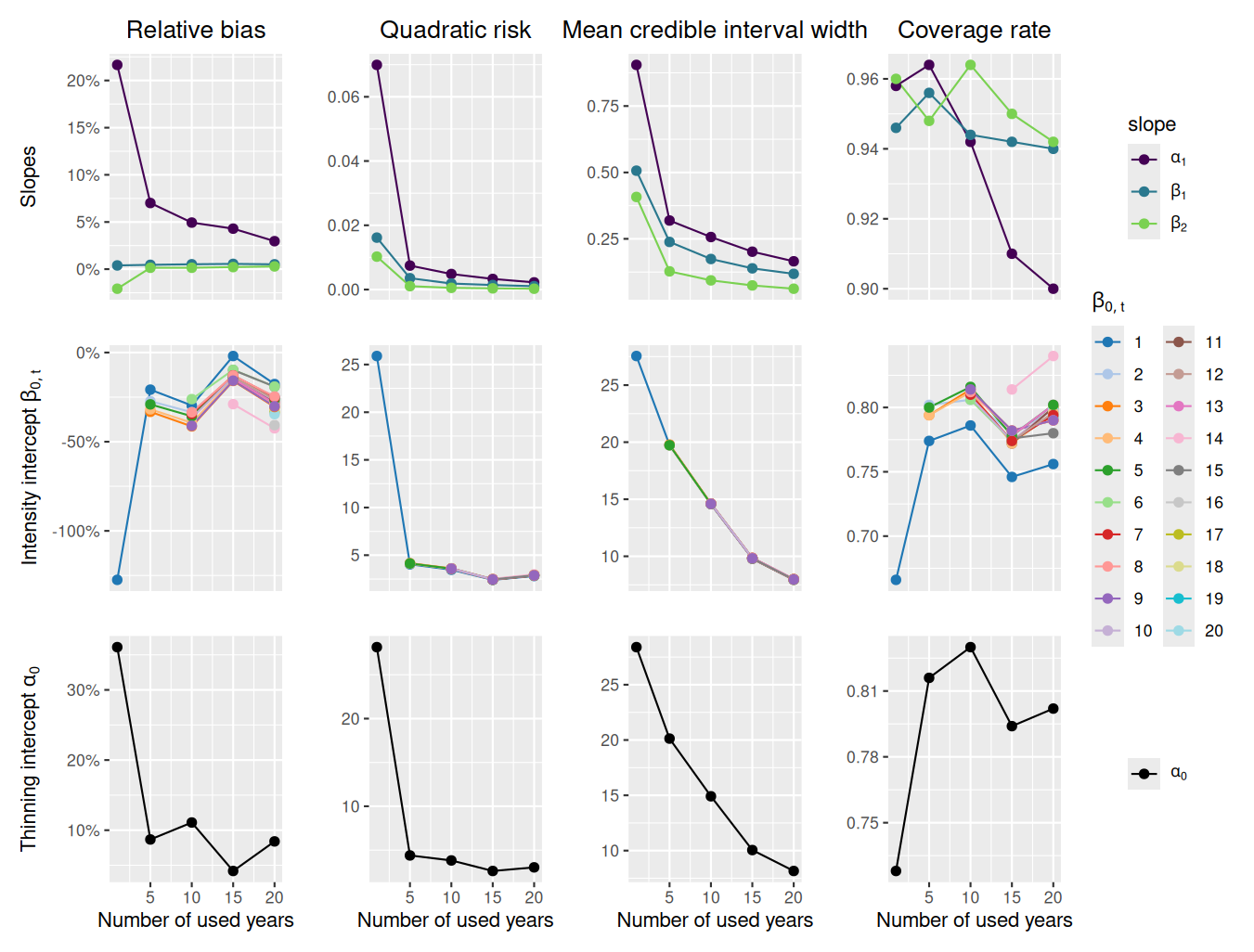}
    \caption{For each coefficient of the $\mathcal M_\text{cst}$ model and each time period considered, from left to right: relative bias, quadratic risk, mean credible interval width, and coverage rate.}
    \label{fig:scen1_cst}
\end{figure}

Figure \ref{fig:scen1_cst} summarizes the performance of the spatio-temporal thinned Poisson point process model with intensity intercepts that are independent across years and constant thinning intercepts, under simulation scenario 1. These results are very similar to those obtained with the $\mathcal M_\text{eff}$ model under this scenario and lead to the same interpretation.

\newpage

\section{Sensitivity analysis of GBIF observation cutoff}\label{annexe:sensib}

\begin{figure}[ht]
\begin{subfigure}{\textwidth}
    \centering
    \includegraphics[width=0.7\linewidth]{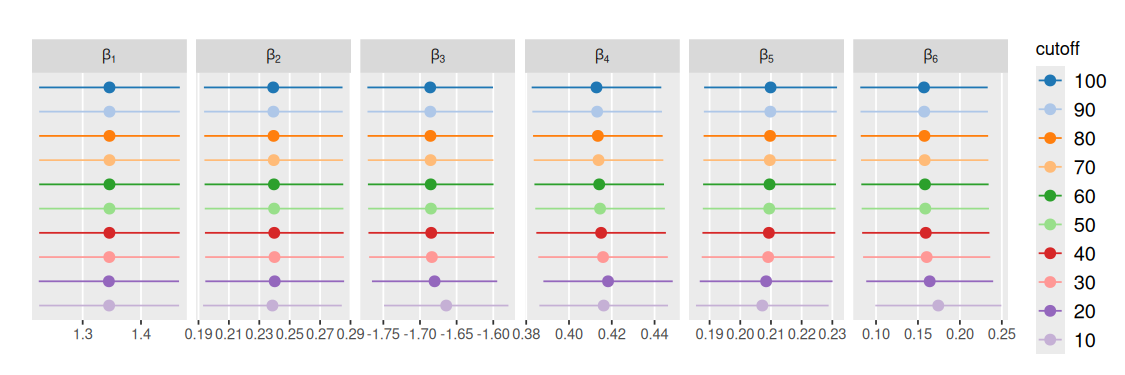}
    \caption{Slopes}
    \label{fig:tronc_pentes}
\end{subfigure}
\vfill
\begin{subfigure}{\textwidth}
    \centering
    \includegraphics[width=0.8\linewidth]{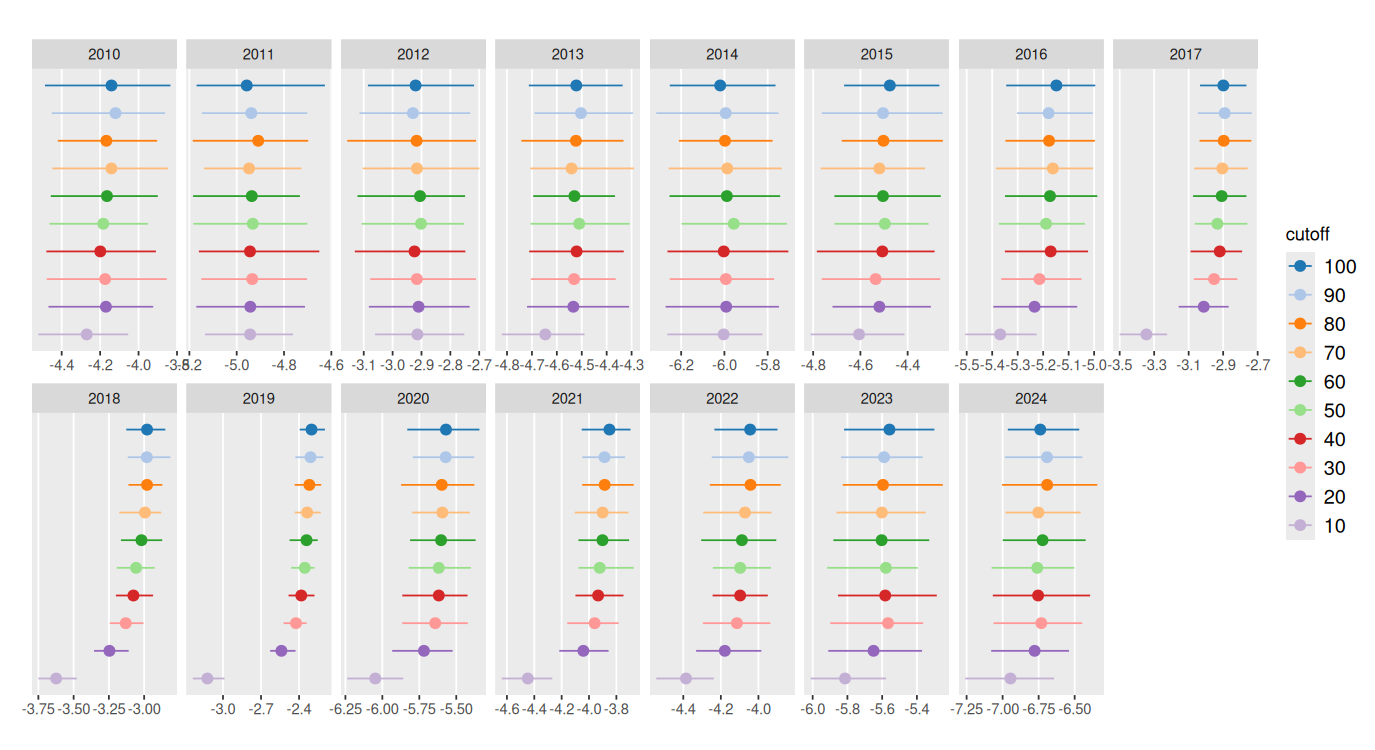}
    \caption{Intensity intercepts $\beta_{0,t}$}
    \label{fig:tronc_intercepts}
\end{subfigure}
\caption{Parameters obtained from the spatio-temporal thinned Poisson point process model $\mathcal M_\text{eff}$ as a function of the cutoff used to construct the thinning covariate $w_{gbif}$.}
\label{fig:tronc}
\end{figure}

The thinning covariate $w_{gbif}$ used in the case study was constructed as the density of observations, across all species, within a 1 km radius of the locations under consideration. In the analysis, the number of observations was capped at a maximum of 50. Here, we assess the sensitivity of the results to the observation cutoff by examining the coefficients of the intensity $\lambda$, which is the main quantity of interest in the model. We fitted the model $\mathcal M_\text{eff}$ for cutoff values  ranging from 10 to 100, in increments of 10 (Figure \ref{fig:tronc}).

The model slopes are extremely similar across the different cutoff values. Only a very slight variation is observed for the smallest cutoff values, particularly for a cutoff of 10. Similar variations are observed for the intensity intercepts, with more pronounced differences (although for larger coefficients).
% and smaller intercepts for the smallest cutoff values.
Overall, we therefore consider that the choice of the cutoff value has little influence on the results. This is also reflected in the WAIC values, which range from 62454 to 62473 and are therefore very similar, except for the cutoff of 10, which yields a WAIC of 62608, corresponding to a substantially larger difference. 
% Thus, the model with a cutoff of 10, which produces different coefficients and consequently a different intensity, provides a poorer fit than the other models and should be discarded.
Finally, this sensitivity analysis supports our choice of a cutoff of 50, with little impact relative to cutoff values between 20 and 100.

\newpage

\section{Comparison of the Poisson point process models fitted to the case study.}\label{annexe_comparaison}

\begin{figure}[ht]
    \centering
    \includegraphics[width=0.8\linewidth]{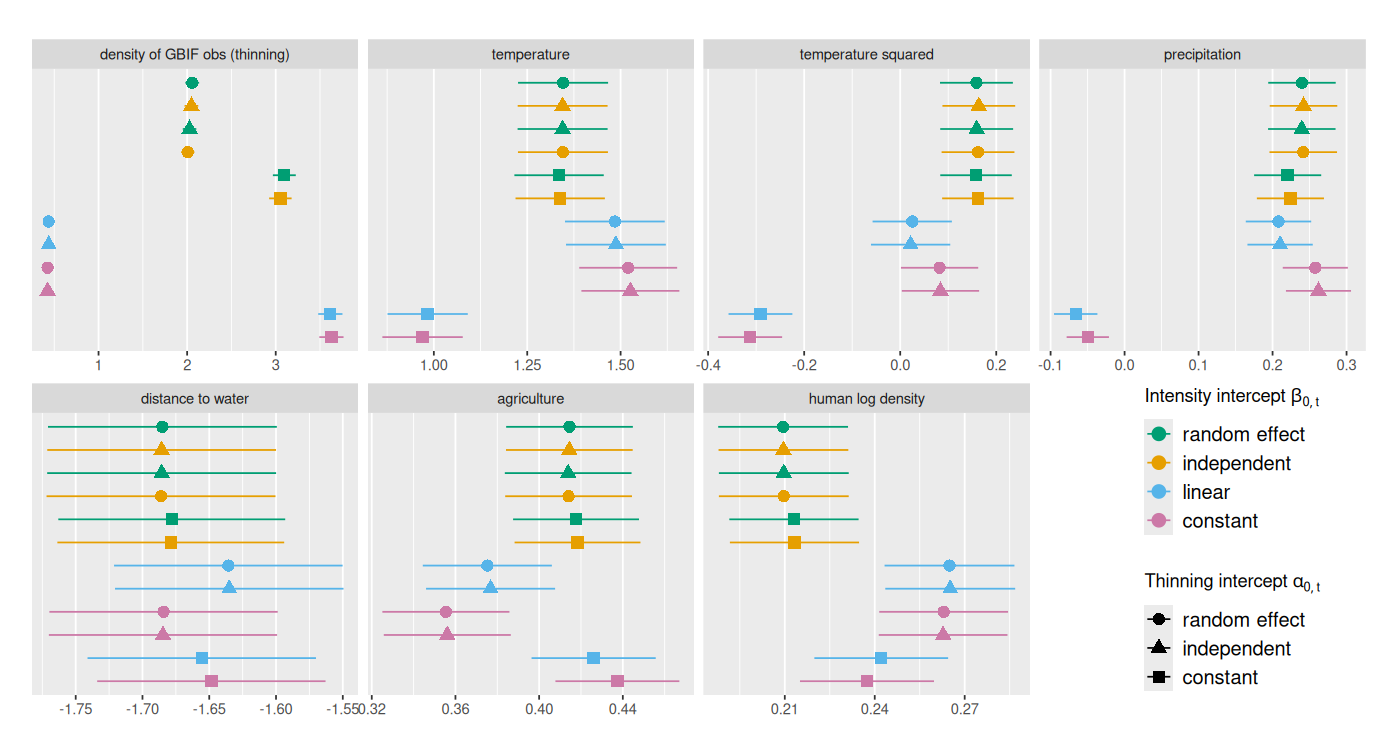}
    \caption{Slopes coefficients for the different Poisson point process models fitted with a quadratic temperature effect. The models are ordered by increasing WAIC, from the best-supported model (top) to the least-supported model (bottom). }
    \label{fig:comp_pentes}
\end{figure}

\begin{figure}
\begin{subfigure}{\textwidth}
    \centering
    \includegraphics[width=0.75\linewidth]{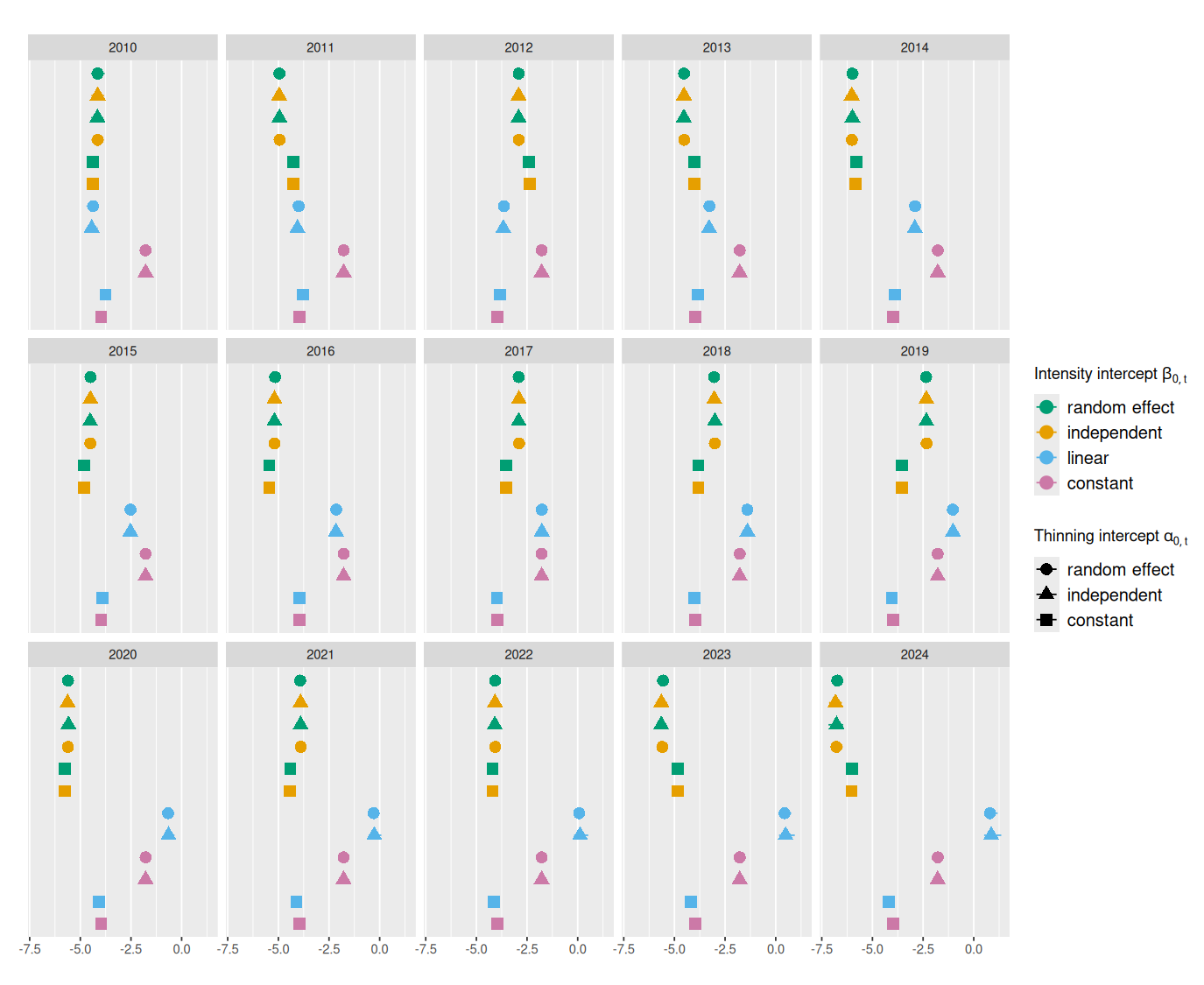}
    \caption{Intensity intercepts}
    \label{fig:comp_beta0}
\end{subfigure}

\begin{subfigure}{\textwidth}
    \centering
    \includegraphics[width=0.75\linewidth]{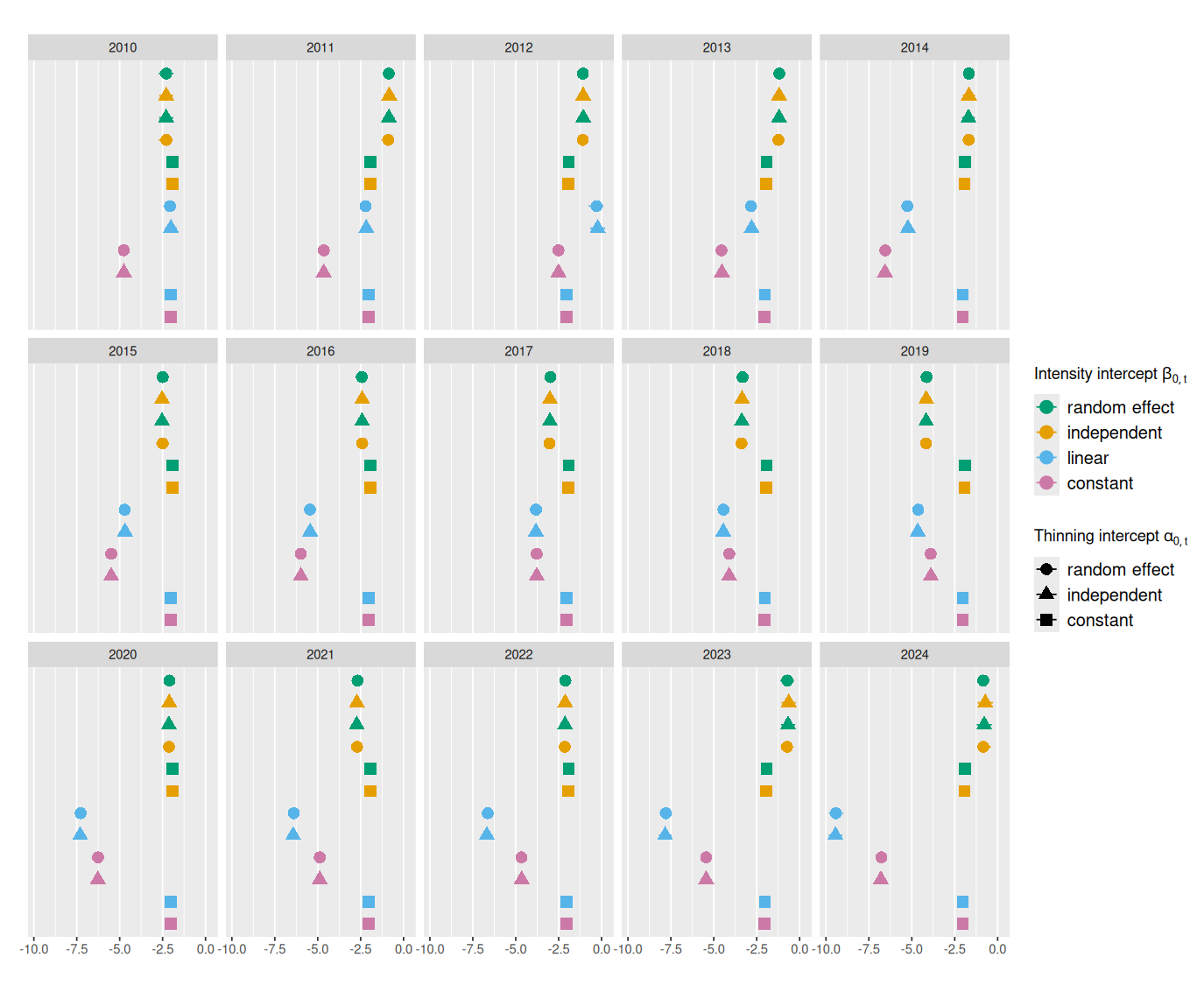}
    \caption{Thinning intercepts}
    \label{fig:comp_alpha0}
\end{subfigure}
\caption{Estimated intercepts for the different Poisson point process models fitted with a quadratic temperature effect. The models are ordered by increasing WAIC, from the model best-supported by the data (top) to the least-supported model (bottom).}
\label{fig:comp}
\end{figure}

% Table \ref{tab:WAIC} compares the different spatio-temporal thinned Poisson point process models fitted to the coypu observation data using the WAIC criterion. In almost all cases, models including a quadratic effect of temperature outperform those including only a linear effect. 
Figures \ref{fig:comp_pentes} and \ref{fig:comp}  compare the estimated coefficients of the different models with a quadratic temperature effect.
The coefficients of the four best-supported models, obtained using either independent or random intercepts for both intercept terms, are estimated almost identically, consistent with their very similar WAIC values.
The intensity coefficients are generally similar across all models. The only noticeable differences concern the coefficients associated with the quadratic temperature and precipitation covariates, whose estimated signs differ between models, although their values remain very close to zero.

\newpage

\section{Estimated log-intensity over the period 2010 - 2024}\label{annexe:carte_periode}

\begin{figure}[ht]
    \centering
    \includegraphics[width=.9\linewidth]{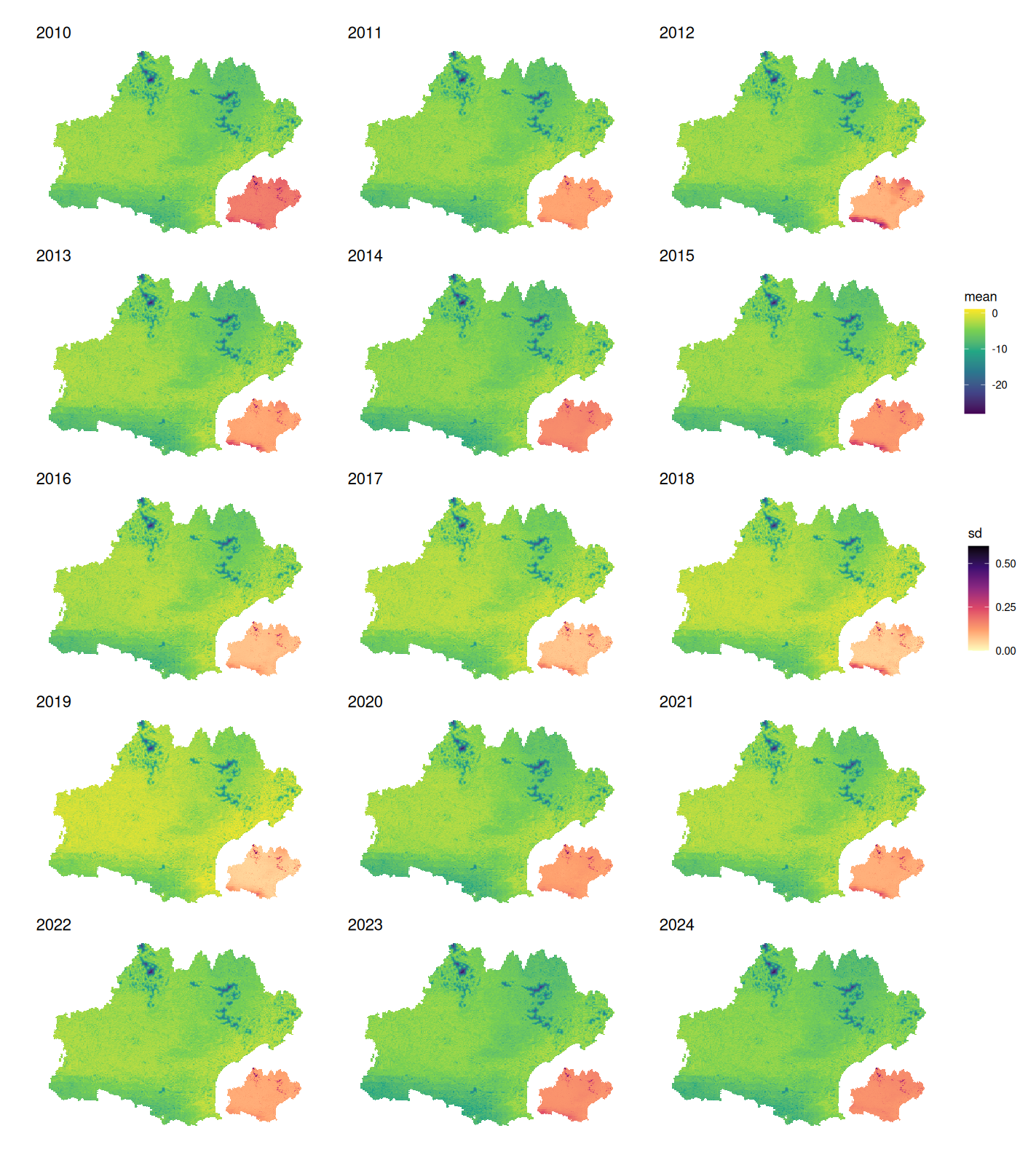}
    \caption{Mean and standard deviation of the estimated log-intensity of the spatio-temporal thinned Poisson point process fitted with two random intercepts. 
    }
    \label{fig:carte_periode}
\end{figure}

\newpage

\section{Validation of the thinned Poisson point process model}\label{annexe:valid}

\begin{figure}
    \centering
    \includegraphics[width=0.9\linewidth]{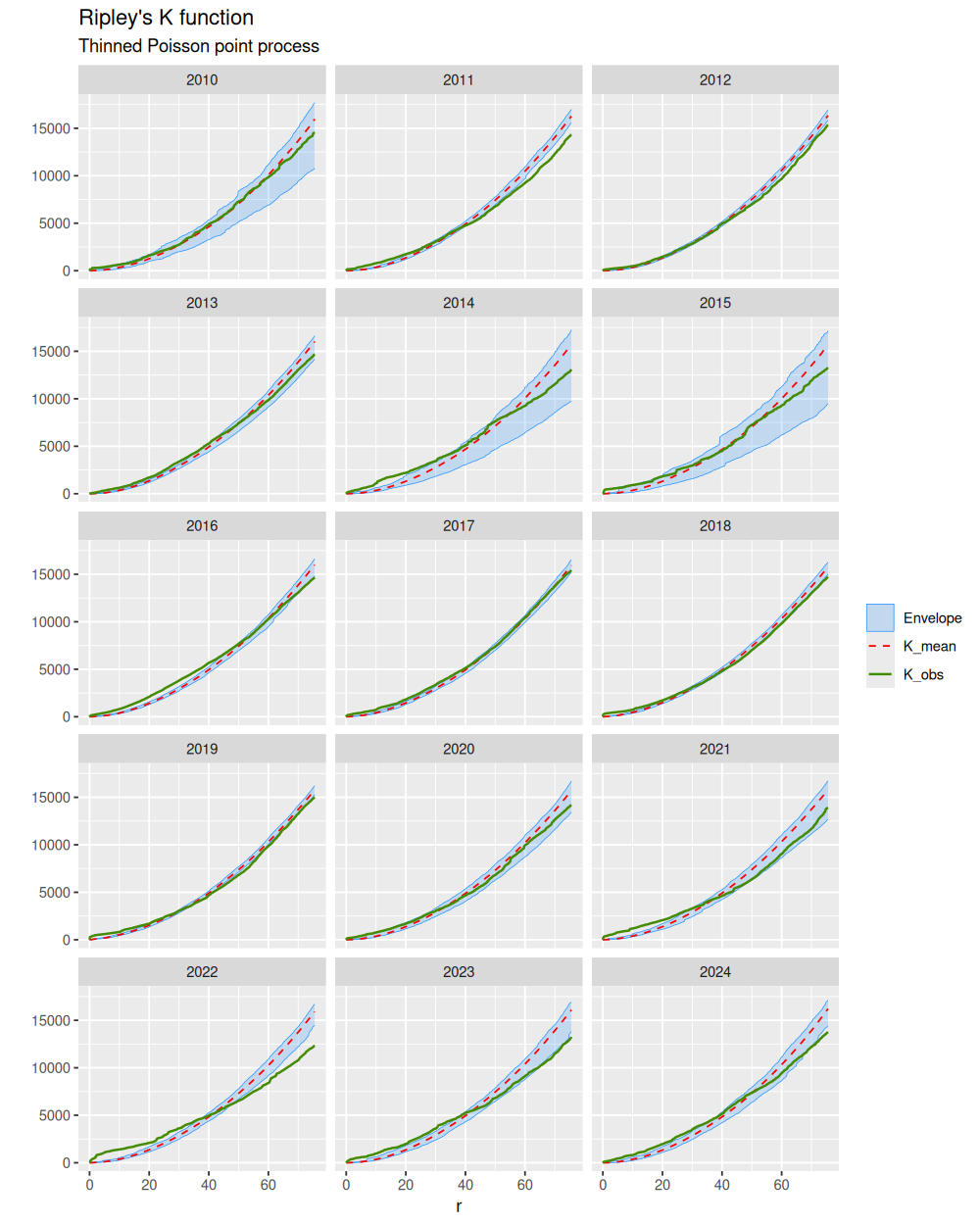}
    \caption{Estimated inhomogeneous Ripley’s $K$-function based on the observed data (solid green line) and on 100 simulations from the fitted thinned point process under the $\mathcal M_\text{eff}$ model (blue envelope). The mean across simulations is shown by the red dashed line. }
    \label{fig:validation}
\end{figure}

We consider here the spatio-temporal thinned Poisson point process model with annual random-effect intercepts ($\mathcal M_\text{eff}$).
To assess the fit of this model, we use Ripley's $K$-function \parencite{ripley_modelling_1977}, in its extension to inhomogeneous point processes introduced by \textcite{baddeley_non_2000}.
% For a Poisson point process $\mathbf{X}$ with intensity $\lambda$ defined on $D$, this function is defined as follows:
% $$K_\text{inhom}(r)=\dfrac 1{|D|}\mathbb E\left[\dsum{s\in \mathbf{X}}{}~\dsum{s'\in\mathbf{X}\setminus\{s\}}{}\dfrac{\mathbb 1(|\!|s-s'|\!|\leqslant r)}{\lambda(s)\lambda(s')}\right], ~r\geqslant 0$$
% where $|D|$ denotes the area of the study domain. This function can be interpreted as a measure of the average clustering of the process at distance $r$. It is a tool for analyzing the point process, although it does not fully characterize the process. An estimator of this function is given by:
% $$\widehat K_\text{inhom}(r)=\dfrac 1{|D|}\dsum{s\in \mathbf{X}}{}~\dsum{s'\in\mathbf{X}\setminus\{s\}}{}\dfrac{\mathbb 1(|\!|s-s'|\!|\leqslant r)}{\lambda(s)\lambda(s')}, ~r\geqslant 0.$$
% Since our study relies solely on observation data, we assess the fit of the thinned Poisson point process model by comparing the estimate of the $K$-function obtained from the observed data with those obtained from data simulated from the fitted model, separately for each year. Figure \ref{fig:validation} presents the results, with the envelopes constructed from 100 simulations. 
Figure \ref{fig:validation} compares, for each year, the $K$-function estimated from the observed data with those estimated from simulations of the fitted thinning process.

We observe that the curve estimated from the observed data consistently follows the trend obtained from the simulations, which is encouraging, but  does not always lie within the envelope. This is particularly noticeable for the year 2022. The observed data exhibit stronger small-scale clustering than the simulated data, together with weaker clustering at larger scales. This is consistent with the finding from the LGCP model (\ref{annexe:lgcp_2}) and indicates that the covariates used in the model fail to capture the small-scale clustering.
However, it is important to emphasize that this analysis assesses only the fit of the observed thinned point process, rather than that of the underlying process of true presences of interest.

\newpage

\section{Extension to a thinned log-Gaussian Cox point process}\label{annexe:lgcp_2}

\subsection{Model}

\begin{figure}[ht]
    \centering
    \includegraphics[width=0.5\linewidth]{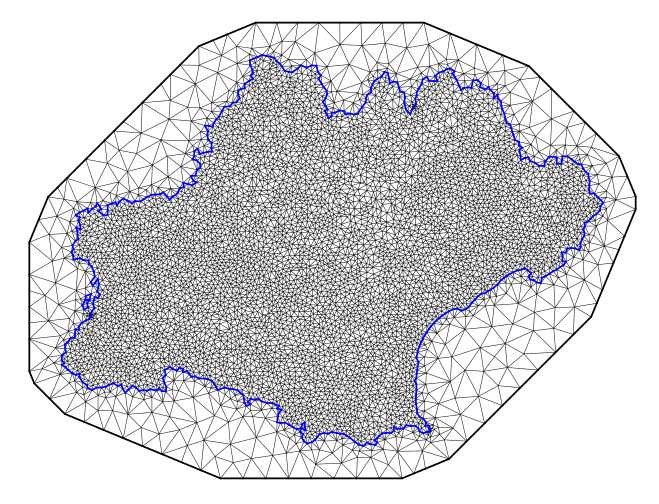}
    \caption{Mesh used for fitting the Poisson and LGCP models using INLA. }
    \label{fig:mesh}
\end{figure}

Using a Poisson point process in a species distribution model relies on the assumption that species presences are mutually independent once the intensity has been specified. Spatial dependence is therefore captured solely through the covariates included in the intensity function. A log-gaussian Cox process (LGCP) relaxes this assumption by considering a Gaussian random log-intensity:
\[\mathbf{X}~|~\boldsymbol{\lambda}\sim PPP(\boldsymbol{\lambda})~~\text{with}~\log(\boldsymbol{\lambda})\text{ a Gaussian field}.\]
The main advantage of this class of Cox processes is that the Gaussian random field can be fitted efficiently using the INLA framework. 
As for Poisson point processes, a Cox process can be thinned with an observation probability $p$ (deterministic in this study) and remains a Cox process, with random intensity $\boldsymbol{\lambda}p$. For further details on Cox processes, we refer the reader to \cite{moller_statistical_2004}. 
As before, $\boldsymbol{\lambda}$ and $p$ can be linked to covariates and the model can be extended to a spatio-temporal setting. In the case study, we fitted a spatio-temporal thinned LGCP over the period 2010–2024, with  intensity and thinning functions defined as follows:
\[\begin{array}{lc}
\log(\boldsymbol{\lambda}_t(s))=&\beta_{0,t}
+\beta_{1}\,x_{temp}(s,t)
+\beta_{2}\,x_{temp^2}(s,t)
+\beta_{3}\,x_{prec}(s,t)
+\beta_{4}\,x_{water}(s)\\
&+\beta_{5}\,x_{agri}(s)
+\beta_{6}\,x_{pop}(s)
+\mathbf Z_t(s)
\end{array}\]
\[\text{logit}(p_t(s))= \alpha_{0,t}+\alpha_1\,w_{gbif}(s,t)\]
with annual random-effect intercepts $\beta_{0,t}$ and $\alpha_{0,t}$, and yearly independent Gaussian random fields $\mathbf Z_t$, with a Matérn covariance function sharing the same hyperparameters for all years. The priors assigned to the model coefficients are the same as those used for the Poisson model. The hyperparameters of the Matérn covariance are assigned penalized complexity (PC) priors satisfying $P(\rho<1)=0.01$ for the range parameter and $P(\sigma>2)=0.01$ for the marginal standard deviation (see \cite{gomez-rubio_chapter_nodate-2} for details on this parametrization). Model fitting was performed using the \textit{inlabru} package, based on the same integration points as those used for the Poisson point process model and the triangular mesh shown in Figure \ref{fig:mesh}.

\subsection{Results}

% résultats illustrés : modèle avec constr = TRUE pour le spde
% très similaires avec ou sans constr, sauf WAIC, largement meilleur avec constr = TRUE

\begin{figure}[ht]
    \centering
    \includegraphics[width=0.9\linewidth]{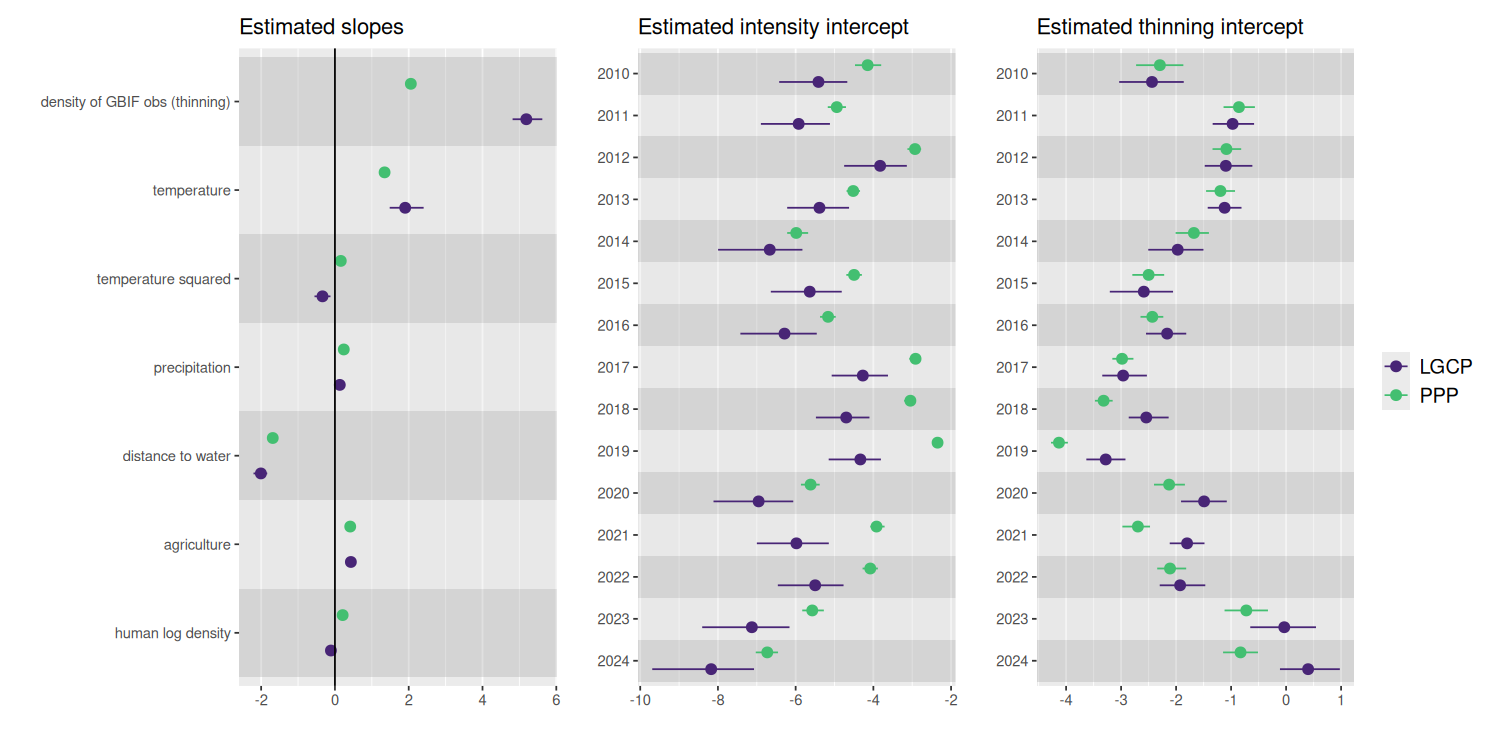}
    \caption{Comparison of the parameter estimates obtained from the spatio-temporal thinned LGCP with those obtained from the spatio-temporal thinned Poisson point process model with random-effect intercepts.}
    \label{fig:comp_PPP_LGCP}
\end{figure}

Figure \ref{fig:comp_PPP_LGCP}  compares parameter estimates from the LGCP and Poisson point process models. Overall, the estimated slopes are broadly consistent between the two models. The main differences concern the slope associated with the thinning covariate, the quadratic temperature and the human density. For these latter two covariates, the estimated slopes have opposite signs across models, although their magnitudes remain close to zero. 
% Apart from these two covariates, the effects of the covariates therefore appear to be similarly identified by the two modelling approaches.
The intercepts also exhibit broadly similar temporal patterns across the two models. However, the intensity intercepts estimated under the LGCP are systematically lower than those obtained under the PPP. A smaller discrepancy in the opposite direction is observed for the thinning intercepts, although this difference is only apparent in the most recent years.
Finally, the LGCP estimates generally exhibit greater uncertainty than their PPP counterparts, with substantially wider credible intervals under the LGCP. 
% This suggests that accounting for the latent Gaussian field introduces additional uncertainty into the estimation of the model parameters.

% Regarding the estimated Gaussian field, the estimates of the two Matérn hyperparameters are reported in Table \ref{tab:hyperpar}, while maps of the posterior mean and standard deviation of the Gaussian field are shown in Figure \ref{fig:GF}.

% \begin{table}[ht]
%     \centering
%     \begin{tabular}{|c|c|c|c|c|c|}
%     \hline
%     Hyperparameter & mean & sd & $2.5\%$ quantile & median & $97.5\%$ quantile\\\hline
%     Range $\rho$ & 16.84 & 0.59 & 15.71 & 16.83 & 18.04\\\hline
%     Standard deviation $\sigma$ & 1.82 & 0.04 & 1.75 & 1.82 & 1.90 \\\hline
%     \end{tabular}
%     \caption{Estimated Matérn covariance hyperparameters of the Gaussian field in the spatio-temporal LGCP model.}
%     \label{tab:hyperpar}
% \end{table}

\begin{figure}[ht]
    \centering
    \includegraphics[width=0.4\linewidth]{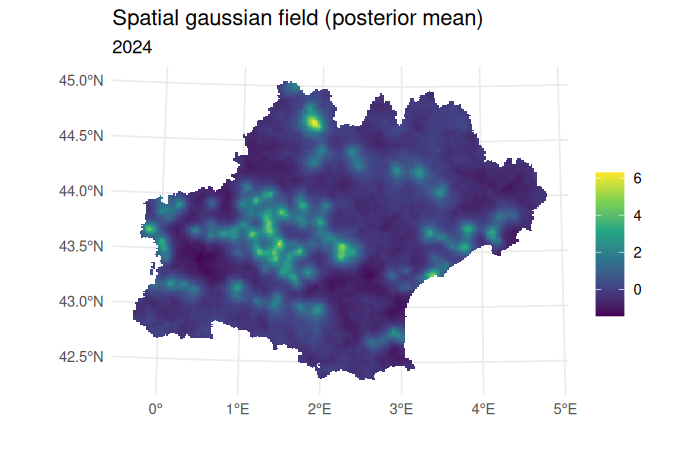}
    \includegraphics[width=0.4\linewidth]{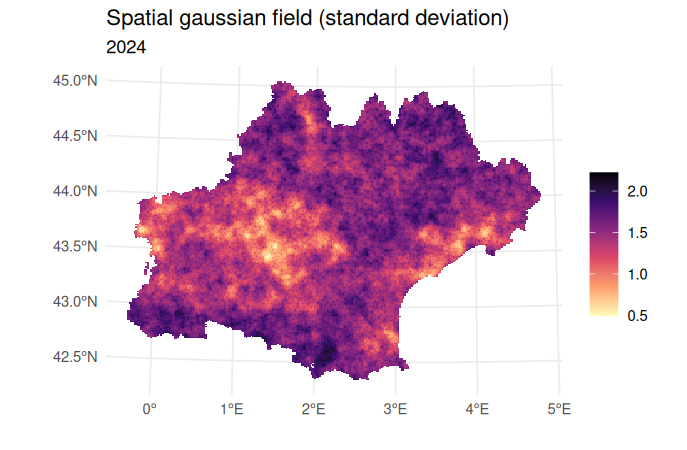}
    \caption{Estimated Gaussian field (posterior mean) and associated standard deviation under the LGCP model. }
    \label{fig:GF}
\end{figure}

The maps of the posterior mean and standard deviation of the Gaussian field are shown in Figure \ref{fig:GF}.
The estimated Gaussian field appears to be rather patchy, with a estimated range of approximately 17 km. This suggests that the covariates included in the model may be insufficient to explain short-range spatial correlations, with the Gaussian field capturing the residual component of this fine-scale correlation. However, we note that the patches of the Gaussian field are spatially close to the observed data. This raises the question of whether the Gaussian field may be confounding the presence intensity $\boldsymbol{\lambda}$ with the thinned observation intensity $\boldsymbol{\lambda} p$.

\begin{figure}[ht]
    \centering
    \includegraphics[width=0.4\linewidth]{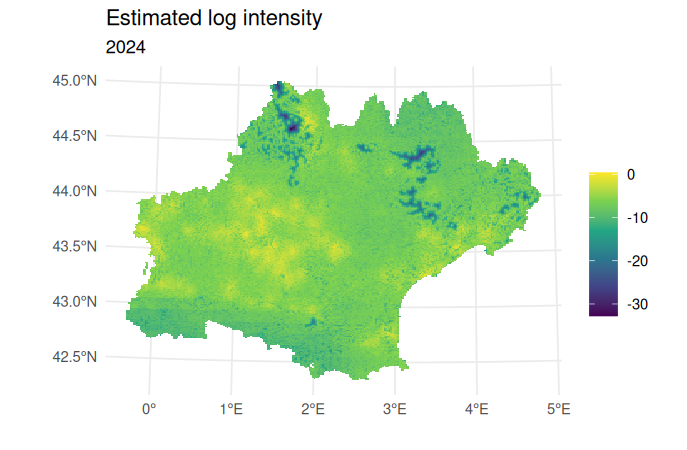}
    \includegraphics[width=0.4\linewidth]{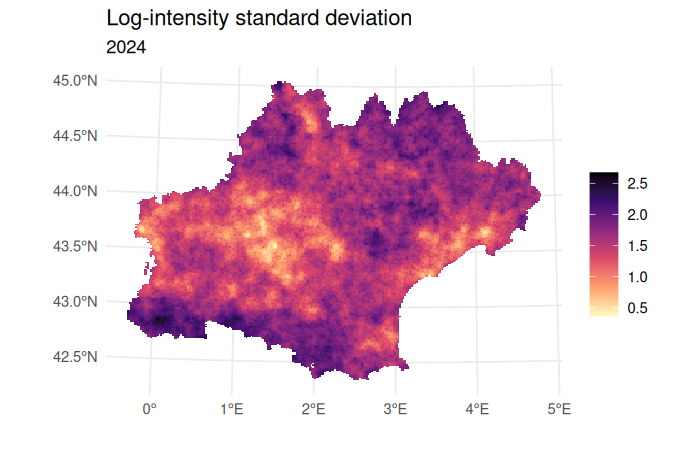}
    \caption{Estimated intensity (posterior mean) and associated standard deviation under the LGCP model.}
    \label{fig:intensité_GF}
\end{figure}

Figure \ref{fig:intensité_GF} shows the intensity estimated from the Gaussian field for the year 2024. The resulting intensity map is fairly similar to that obtained with the Poisson point process. However, the high-intensity areas are more localized and closer to the observed data. In addition, the standard deviations are substantially larger under this second model, and the WAIC is markedly worse (66020.77 compared with 62460.81 for the Poisson model). 

\printbibliography[title={References}]

\end{document}